\documentclass[twocolumn, twoside]{article}
\usepackage{geometry} 
\usepackage[linktoc=all]{hyperref}
\hypersetup{
	bookmarksopen=true,
	bookmarksdepth=3,
	colorlinks=true,
	citecolor=blue,
	linkcolor=[rgb]{0.6,0,0}
}
\usepackage{graphicx}
\usepackage{cuted}
\usepackage{subcaption}
\usepackage{amssymb,amsmath}
\usepackage{listings}
\usepackage{xcolor}
\newcommand{\scalar}{\cdot}
\newcommand{\cross}{\times}

\newcommand{\degree}{^{\circ}}
\newcommand{\at}[2]{\left.#1\,\right|_{_{#2}}}

\newcommand{\atan}{\text{atan}}

\newcommand{\markedsection}[2]{\section[#1]{#1%
\sectionmark{#2}}
\sectionmark{#2}}
\newcommand{\orcid}[1]{\href{https://orcid.org/#1}{\includegraphics[width=8pt]{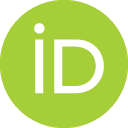}}}

\usepackage[style=numeric-comp, sorting=none, backend=bibtex]{biblatex}
\bibliography{bibliography}

\usepackage{fancyhdr}
\begin{document}

\title{Analytic Solution to the Piecewise Linear Interface Construction Problem and its Application in Curvature Calculation for Volume-of-Fluid Simulation Codes}
\author{Moritz Lehmann \orcid{0000-0002-4652-8383}\ \ \ \ Stephan Gekle \orcid{0000-0001-5597-1160}\\
 Biofluid Simulation and Modeling -- Theoretische Physik VI\\University of Bayreuth, 95440 Bayreuth, Germany\\
              Tel.: +49 921 55 3094, eMail: moritz.lehmann@uni-bayreuth.de
}
\date{February 18, 2021}
\maketitle
\section*{Keywords}
PLIC, plane-cube intersection, Volume-of-Fluid, Lattice Boltzmann Method, GPU, OpenCL


\section*{Abstract}
The plane-cube intersection problem has been around in literature since 1984 and iterative solutions to it have been used as part of piecewise linear interface construction (PLIC) in computational fluid dynamics simulation codes ever since. 
In many cases, PLIC is the bottleneck of these simulations regarding compute time, so a faster, analytic solution to the plane-cube intersection would greatly reduce compute time for such simulations. 
We derive an analytic solution for all intersection cases and compare it to the one previous solution from Scardovelli and Zaleski (Ruben Scardovelli and Stephane Zaleski. "Analytical relations connecting linear interfaces and
volume fractions in rectangular grids". In: Journal of Computational Physics 164.1 (2000), pp. 228–
237.), which we further improve to include edge cases and micro-optimize to reduce arithmetic operations and branching.
We then extend our comparison regarding compute time and accuracy to include two different iterative solutions as well.
We find that the best choice depends on the employed hardware platform: on the CPU, Newton-Raphson is fastest with vectorization while analytic solutions perform better without. The reason for this is that vectorization instruction sets do not include trigonometric functions as used in the analytic solutions. 
On the GPU, the fastest method is our optimized version of the analytic SZ solution.\\
We finally provide details on one of the applications of PLIC -- curvature calculation for the Volume-of-Fluid model used for free surface fluid simulations in combination with the lattice Boltzmann method.

\section{Introduction}
Piecewise linear interface construction (PLIC) -- first occurring in literature for 2D in 1982 \cite{youngs1982time} and for 3D in 1984 \cite{youngs1984interface} -- refers to the problem of calculating the offset along the given normal vector of a plane intersecting a unit cube for a given truncated volume. There are five possible intersection cases (cf. figure \ref{fig:plic-cases}), of which the numbers (1), (2) and (5) have been already solved in the original 1984 work by Youngs \cite{youngs1984interface}, but the cubic polynomial cases (3) and (4) -- resigned as impossible to algebraically invert \cite{janssen2013enhanced} -- in the majority of literature are approximated by a Newton-Raphson iterative solution. 
Nevertheless, there does exist an analytic solution by Scardovelli and Zaleski (SZ) \cite{scardovelli2000analytical} and a single documented implementation thereof in Fortran \cite{kawano2016simple} which also includes an approximative version termed APPLIC.

Here, we formulate the PLIC problem from the ground up -- first in the inverse direction -- and derive an alternative analytic solution for all intersection cases by inverting the inverse formulation. 
We then compare our novel solution with (i) the original SZ solution, (ii) an improved and micro-optimized version of the SZ solution developed in the present work, (iii) an iterative solution using Newton-Raphson and (iv) an iterative solution using nested intervals. 
Depending on the available microarchitecture (GPU/CPU), vectorization may be available, which strongly favors multiplications and additions while not speeding up trigonometric functions, impacting which of the algorithms is fastest.

Among the applications for PLIC are Volume-of-Fluid simulation codes such as \textsl{FluidX3D} \cite{lehmann2019high, haeusl2019mpi} and others \cite{bogner2016curvature, korner2005lattice, thurey2005interactive, pohl2008high, schreiber2010gpu, popinet2009accurate, jafari2007improved, xing2007lattice, donath2011wetting, donath2011verification, anderl2014free}, 
often in conjunction with GPU implementations \cite{lehmann2019high, haeusl2019mpi, obrecht2011new, wittmann2016hardware, delbosc2014optimized, herschlag2018gpu, mawson2014memory, wittmann2013comparison, kuznik2010lbm} of the lattice Boltzmann method \cite{kruger2017lattice, chapman1990mathematical, purqon2017accuracy}, used for simulating free surface fluid flows. 
In particular, these simulations work on a cubic lattice with every lattice point having a fill level assigned to it and PLIC is used in the process of surface reconstruction during curvature calculation for calculating physical surface tension effects \cite{lehmann2019high, bogner2016curvature}. 
In the final section of this work, we provide a detailed overview on the state-of-the-art curvature calculation procedure using PLIC.

\section{Plane-Cube Intersection} \label{sec-plic-cube}
Inputs to the PLIC algorithm are the truncated volume $V_0\in[0,1]$ and the (normalized) normal vector of the plane $\vec{n}=(n_x,n_y,n_z)^ \mathrm{T}$, $|\vec{n}|=n_x^2+n_y^2+n_z^2=1$. 
The desired output is the plane offset from the origin (center of the unit cube) along the normal vector $d_0$
\begin{equation}
V_0,\,(n_x,n_y,n_z)^ \mathrm{T}\ \to\ d_0
\end{equation}
where $d_0\in[-\frac{|n_x|+|n_y|+|n_z|}{2},\frac{|n_x|+|n_y|+|n_z|}{2}]$. 
The interval is determined by the normal vector orientation: depending on the normal vector, the maximum possible distance from the cube center to be still at least touching the cube in one point varies between $\frac{1}{2}$ (normal vector parallel to one of the coordinate system axes) and $\frac{\sqrt{3}}{2}$ (normal vector along the space diagonal). 

\subsection{Applying symmetry conditions to reduce problem complexity}
\begin{figure*}[!htbp]
\centering \includegraphics[width=\textwidth]{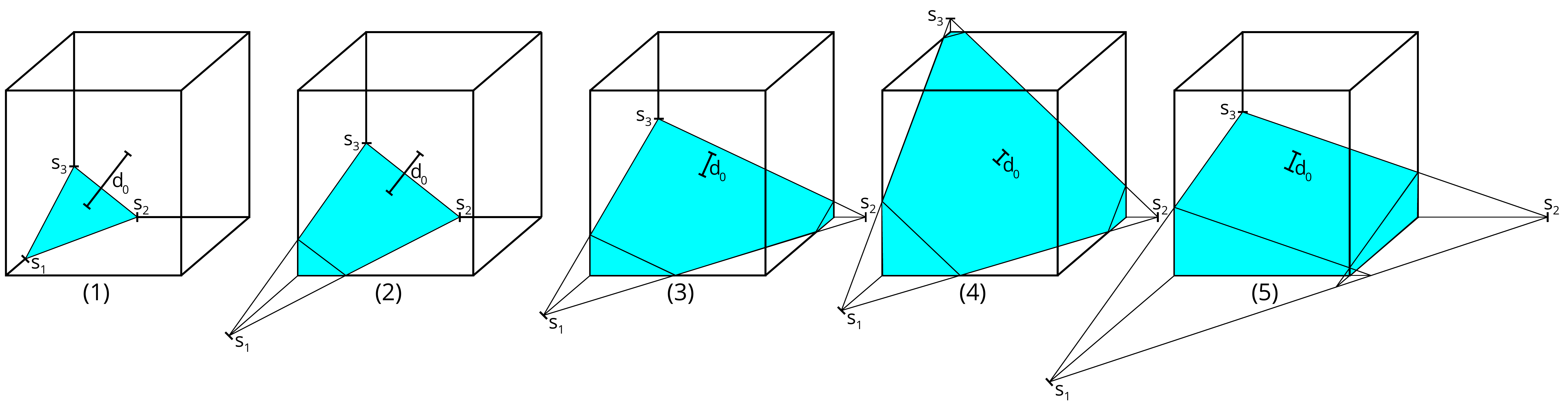}
\caption{All possible intersection cases of a plane and a unit cube. The truncated volume of cases (1) to (4) is a tetrahedral pyramid with zero (1), one (2), two (3) or all three (4) corners extending outside of the unit cube being cut-off tetrahedral pyramids themselves. 
} 
\label{fig:plic-cases}
\end{figure*}

To reduce the amount of possible cases and to avoid having to consider all possible intersections of the plane and cube edges -- following the scheme in \cite{youngs1984interface} and \cite{jafari2007improved}\footnote{We note that in \cite{jafari2007improved} in equations (21) and (23) respectively the "$+$" should be a "$-$" and in equation (24) the "$>$" should be a "$<$".} -- the normal vector is component-wise mirrored into positive. The mirrored normal vector components are sorted ascending for their magnitude such that $0\leq n_1\leq n_2 \leq n_3\leq1$. 
Because $\vec{n}$ is normalized, the absolute value of its largest component $n_3$ is always greater than zero. 
\begin{align} \label{eq-plic-sort-1}
n_1&:=\min(|n_x|,\,|n_y|,\,|n_z|)\geq0\\
n_3&:=\max(|n_x|,\,|n_y|,\,|n_z|)>0\\ \label{eq-plic-sort-3}
n_2&:=|n_x|+|n_y|+|n_z|-n_1-n_3\geq0
\end{align}
Since the function $V_0(d_0)$ is symmetric around $d_0=0$ and increasing monotonically, the reduced-symmetry-volume $V\in[0,\frac{1}{2}]$ is limited to the lower half of the intersection volume $V_0$ and the upper half is reconstructed from symmetry.
\begin{align} \label{eq-plic-V}
V:=&\ \frac{1}{2}-\left|V_0-\frac{1}{2}\right|\\ \label{eq-plic-V0}
V_0=&\ \text{sign}(d_0)\left(\frac{1}{2}-V\right)+\frac{1}{2}
\end{align}
This symmetry condition for the case $V_0>\frac{1}{2}$ is now applied to $d_0$, and the coordinate origin is shifted from $(0,\,0,\,0)$ (center of the unit cube) to $(-\frac{1}{2},\,-\frac{1}{2},\,-\frac{1}{2})$ (bottom left corner in the back in fig. \ref{fig:plic-cases}), 
resulting in the distance $d\in[0,\frac{n_1+n_2+n_3}{2}]$ in reduced symmetry space:
\begin{align} \label{eq-plic-d}
d:=&\ \frac{n_1+n_2+n_3}{2}-|d_0|\\ \label{eq-plic-d0}
d_0=&\ \text{sign}\left(V_0-\frac{1}{2}\right)\left(\frac{n_1+n_2+n_3}{2}-d\right)
\end{align}
With this reduction in symmetry, there are only five different intersection cases remaining (see figure \ref{fig:plic-cases}).

\newpage
\subsection{Formulating the inverse PLIC problem}
In order to derive the analytic PLIC solution, first the inverse problem is formulated in equations -- again following the scheme in \cite{youngs1984interface}. 
In the inverse problem, the intersection volume is calculated from the plane offset and normal vector as inputs. 
At first, the intersection points $s_1$, $s_2$ and $s_3$ of the plane with the coordinate system axes (see figure~\ref{fig:plic-cases}) are determined:
\begin{equation}
s_1:=\frac{d}{n_1}\ \ \ \geq\ \ \ s_2:=\frac{d}{n_2}\ \ \ \geq\ \ \ s_3:=\frac{d}{n_3}
\end{equation}
Now one calculates the actual volume in reduced symmetry space. 
The approach is to calculate the volume of the tetrahedral pyramid formed by the plane and the coordinate system axes and, if necessary, subtract the volumes of zero, one, two or all three corners that extend beyond $1$. For case (3), an additional condition is required to mutually exclude case (5), in which the bottom four corners of the cube are located beneath the plane. 
\newpage
\begin{strip}
\begin{equation} \label{eq-plic-cube-volume-1}
V=
\begin{cases}
\frac{1}{6}\,s_1\,s_2\,s_3\ \ \ \ \ \ \ \ \ \ \ \ \ \ \ \ \ \ \ \ \ \ \ \ \ \ \ \ \ \ \ \ \ \ \ \ \ \ \ \ \ \ \ \ \ \ \ \ \ \ \ \ \ \ \ \ \ \ \ \ \ \ \ \ \ \ \ \,\text{ if }\ \ \ \ \ \ \ \ \ \ \ \ \ \ \ s_1<1\\
\frac{1}{6}\,s_2\,s_3\left(s_1-(s_1-1)\left(1-\frac{1}{s_1}\right)^2\right)\ \ \ \ \ \ \ \ \ \ \ \ \ \ \ \ \ \ \ \ \ \ \ \ \ \ \ \ \ \ \ \ \ \ \ \text{ if }s_1\geq1\text{ and }s_2\leq1\\
\frac{1}{6}\,s_3\left(s_1\,s_2-(s_1-1)\,s_2\left(1-\frac{1}{s_1}\right)^2-(s_2-1)\,s_1\left(1-\frac{1}{s_2}\right)^2\right) \ \text{ if }s_2\geq1\text{ and }s_3\leq1\text{ and }s_1\,(s_2-1)\leq s_2\\
\frac{1}{6}\left(s_1\,s_2\,s_3-(s_1-1)\,s_2\,s_3\left(1-\frac{1}{s_1}\right)^2-(s_2-1)\,s_1\,s_3\left(1-\frac{1}{s_2}\right)^2-(s_3-1)\,s_1\,s_2\left(1-\frac{1}{s_3}\right)^2\right)\ \ \text{ if }s_3\geq1\\
\frac{1}{2}\,s_3\left(2-\frac{1}{s_1}-\frac{1}{s_2}\right)\ \ \ \ \ \ \ \ \ \ \ \ \ \ \ \ \ \ \ \ \ \ \ \ \ \ \ \ \ \ \ \ \ \ \ \ \ \ \ \ \ \ \ \ \ \ \ \ \ \ \ \ \ \ \,\text{ otherwise}
\end{cases}
\end{equation}
\end{strip}
\begin{strip}
To shorten equation \eqref{eq-plic-cube-volume-1}, $s_1$, $s_2$ and $s_3$ are substituted and the expression is simplified, yielding
\begin{equation} \label{eq-plic-cube-volume-2}
V=\frac{1}{6\,n_1\,n_2\,n_3}\cdot
\begin{cases}
\ d^3 &\text{ (1) if }\ \ \ \ \ \ \ d<n_1\\
(d^3-(d-n_1)^3) &\text{ (2) if }n_1\leq d\leq n_2\\
(d^3-(d-n_1)^3-(d-n_2)^3) &\text{ (3) if }n_2\leq d\leq\min(n_1+n_2,\,n_3)\\
(d^3-(d-n_1)^3-(d-n_2)^3-(d-n_3)^3) &\text{ (4) if }n_3\leq d\\
\ 6\,n_1\,n_2\,(d-\frac{1}{2}\,(n_1+n_2)) &\text{ (5) if }\min(n_1+n_2, n_3)\leq d\leq n_3
\end{cases}
\end{equation}
\end{strip}
which is already quite a bit more friendly and completes the inverse PLIC formulation in conjunction with equations \eqref{eq-plic-sort-1} to \eqref{eq-plic-sort-3}, \eqref{eq-plic-d} and \eqref{eq-plic-V0}. The condition for case (5) is the remaining free sector of the possible range of $d$ mutually excluded by the other four cases. Listing \ref{list-inverse-plic-implementation} shows the fully optimized OpenCL C implementation of the inverse PLIC solution.
\begin{lstlisting}[caption={Fully optimized OpenCL C implementation of the inverse PLIC solution. To avoid branching between cases (3) and (4), in the implementation the \texttt{fdimf(x,y):=max(x-y,0)} function is used. Case (2) cannot be included with another \texttt{fdimf(x,y)} because in case (2) division by $n_1$ must be avoided since it could be zero.},captionpos=b,label={list-inverse-plic-implementation}]
float plic_cube_inverse(const float d0, const float3 n) { // unit cube - plane intersection: plane offset d0, normal vector n -> volume V0 in [0,1]
	const float n1 = fmin(fmin(fabs(n.x), fabs(n.y)), fabs(n.z)); // eliminate most cases due to symmetry
	const float n3 = fmax(fmax(fabs(n.x), fabs(n.y)), fabs(n.z));
	const float n2 = fabs(n.x)-n1+fabs(n.y)+fabs(n.z)-n3;
	const float d = 0.5f*(n1+n2+n3)-fabs(d0); // calculate PLIC with reduced symmetry, shift origin from (0.0,0.0,0.0) -> (0.5,0.5,0.5)
	float V; // 0.0<=V<=0.5
	if(fmin(n1+n2, n3)<=d && d<=n3) { // case (5)
		V = (d-0.5f*(n1+n2))/n3; // avoid division by n1 and n2
	} else if(d<n1) { // case (1)
		V = cb(d)/(6.0f*n1*n2*n3); // condition d<n1==0 is impossible if d==0.0f
	} else if(d<=n2) { // case (2)
		V = (3.0f*d*(d-n1)+sq(n1))/(6.0f*n2*n3); // avoid division by n1
	} else { // case (3) or (4)
		V = (cb(d)-cb(d-n1)-cb(d-n2)-cb(fdim(d, n3)))/(6.0f*n1*n2*n3);
	}
	return copysign(0.5f-V, d0)+0.5f; // apply symmetry for V0>0.5
}
\end{lstlisting}

\subsection{Inverting the inverse PLIC Formulation analytically} \label{sec-plic}
Equation \eqref{eq-plic-cube-volume-2} is now inverted for each case individually. 
Cases (1), (2) and (5) are easy, but cases (3) and (4) are non-trivial third order polynomials. 
Here we make use of the tool Mathematica, which outputs three complex solutions for cases (3) and (4) each (section \ref{appendix-a}), of which the third solutions respectively are the right ones as their imaginary parts in the desired range are zero after simplification. 
Luckily, both are of the same overall form (eq. \eqref{eq-plic-complex}). 
However, a complex solution is not useful here since the expected result is a real number -- a problem known as the \textit{casus irreducibilis} -- and most programming languages cannot deal with complex numbers natively. It would also lead to unwanted computational overhead to carry along the imaginary part during computation, which in the end will be zero anyway. To overcome this, we again make use of Mathematica to simplify the general form of the complex solution (eq. \eqref{eq-plic-complex}) in order to obtain the real, trigonometric solution, which is then even further simplified using the trigonometric identity $\sqrt{3}\,\sin(\alpha)-\cos(\alpha)=-2\,\sin(\frac{\pi}{6}-\alpha)$ such that the number of trigonometric functions (which are computationally expensive compared to simpler operations such as additions or multiplications) is minimized (eq. \eqref{eq-plic-real}).
\newpage
\begin{strip}
\begin{align} \label{eq-plic-real}
\text{f}\,(x,\,y,\,a,\,b,\,c):=&\,c-2\,\frac{a+b\,\sqrt[3]{x^2+y^2}}{\sqrt[6]{x^2+y^2}}\sin\left(\frac{\pi}{6}-\frac{1}{3}\,\atan2(y,x)\right)=\\ \label{eq-plic-complex}
=&\,c-a\,\frac{(1-i\sqrt{3})}{\sqrt[3]{x+i\,y}}-b\,(1+i\sqrt{3})\sqrt[3]{x+i\,y}
\end{align}
\end{strip}
\noindent For better readability, a few expressions are pre-defined. Hereby the normalization condition $n_1^2+n_2^2+n_3^2=1$ is applied.
\begin{align} \label{eq-plic-x3}
x_3&:=81\,n_1\,n_2\,(n_1+n_2-2\,V\,n_3)>0\\
y_3&:=\sqrt{23328\,(n_1\,n_2)^3-x_3^2}\geq0\\
a_3&:=\sqrt[3]{54}\ n_1\,n_2\\
b_3&:=\frac{1}{\sqrt[3]{432}}\\
c_3&:=n_1+n_2
\end{align}
\begin{align} \label{eq-plic-y3}
t_4&:=9\,(n_1+n_2+n_3)^2-18\\
x_4&:=324\,n_1\,n_2\,n_3\,(1-2\,V)\geq0\\ \label{eq-plic-y4}
y_4&:=\sqrt{4\,t_4^3-x_4^2}\geq0\\
a_4&:=\frac{1}{\sqrt[3]{864}}\,t_4\\
b_4&:=\frac{1}{\sqrt[3]{3456}}\\ \label{eq-plic-c4}
c_4&:=\frac{n_1+n_2+n_3}{2}
\end{align}
Finally then, the complete analytic solution to the 3D PLIC problem is given by\\
\begin{strip}
\begin{equation} \label{eq-plic-cube-offset}
d=
\begin{cases}
d_1=\sqrt[3]{6\,V\,n_1\,n_2\,n_3} &\text{ (1) if }\ \ \ \ \ \ \ d_1<n_1\\
d_2=\frac{n_1}{2}+\sqrt{2\,V\,n_2\,n_3-\frac{1}{12}\,n_1^2} &\text{ (2) if }n_1\leq d_2\leq n_2\\
d_3=\text{f}\,(x_3,\,y_3,\,a_3,\,b_3,\,c_3) &\text{ (3) if }n_2\leq d_3\leq\min(n_1+n_2,\,n_3)\\
d_4=\text{f}\,(x_4,\,y_4,\,a_4,\,b_4,\,c_4) &\text{ (4) if }n_3\leq d_4\\
d_5=V\,n_3+\frac{n_1+n_2}{2} &\text{ (5) if }\min(n_1+n_2, n_3)\leq d_5\leq n_3
\end{cases}
\end{equation}
\end{strip}
in conjunction with equations  \eqref{eq-plic-sort-1} to \eqref{eq-plic-sort-3}, \eqref{eq-plic-V}, \eqref{eq-plic-d0}, \eqref{eq-plic-real} and \eqref{eq-plic-x3} to \eqref{eq-plic-c4}.\\\\
In equation \eqref{eq-plic-cube-offset} it is noteworthy that the conditions for the five different cases are determined a posteriori by the result itself. 
This means that each case has to be evaluated successively and for the resulting value $d$ the respective condition has to be tested. 
If the condition is true, calculation is stopped and $d$ is returned. 
If the condition is false, the next case has to be evaluated and so on, until the last case is reached.

The order in which the cases are computed and checked can be optimized to calculate the most difficult and infrequent cases last, when the probability is high that one of the easier and more frequent cases has already been chosen. 'Frequent' here refers to some cases appearing more often than others with randomized $V_0$ and $\vec{n}$ as expected in typical PLIC applications. 
Here also special considerations for edge cases (more on that below) need to be taken into account to avoid possible divisions by zero. With this in mind, the order (5)$\to$(2)$\to$(1)$\to$(3)$\to$(4) is preferred.

Additional speedup can be gained by noting that the implicit condition involving $d$ can be replaced by an explicit condition involving $V$ for cases (1), (2) and (5):\\
\begin{strip}
\begin{equation} \label{eq-plic-cube-offset-a-priori}
d=
\begin{cases}
d_1=\sqrt[3]{6\,V\,n_1\,n_2\,n_3} &\text{ (1) if }\ \ \ \ \ \ \ \ \ \ \ \ \ \ \ \ \ 6\,V\,n_2\,n_3<n_1^2\\
d_2=\frac{n_1}{2}+\sqrt{2\,V\,n_2\,n_3-\frac{1}{12}\,n_1^2} &\text{ (2) if }\ 3\,n_2\,(V\,n_3+n_1-n_2)\leq n_1^2\leq6\,V\,n_2\,n_3\\
d_3=\text{f}\,(x_3,\,y_3,\,a_3,\,b_3,\,c_3) &\text{ (3) if }\ \ \ \ \ \ \ \ \ \ \ \ \ \ \ n_2\leq d_3\leq\min(n_1+n_2,\,n_3)\\
d_4=\text{f}\,(x_4,\,y_4,\,a_4,\,b_4,\,c_4) &\text{ (4) if }\ \ \ \ \ \ \ \ \ \ \ \ \ \ \ \ \ \ \ \ \ \ \ \ \ \,n_3\leq d_4\\
d_5=V\,n_3+\frac{n_1+n_2}{2} &\text{ (5) if }\ \ \ \ \ \ \ \ \ \ \ \ \ \ \ \ \ \ \ n_1+n_2\leq2\,V\,n_3
\end{cases}
\end{equation}
\end{strip}
Since $V$ is known, these conditions are checked a priori in order to avoid root function calls if the condition is false.

For even more speedup, all redundant mathematical operations are reduced to a minimum by pre-calculating them to variables (micro-optimization) and condition checks mutually excluded by previous checks are skipped, especially all conditions for the very last case. In the implementation order $(5)\to(2)\to(1)\to(3)\to(4)$ the conditions for case (3) simplify to $d_3\leq n_3$, which will mutually exclusive decide between cases (3) and (4). Since both $d_3$ and $d_4$ are very complicated expressions, here no simplified a priori condition is formulated. 

In equations \eqref{eq-plic-y3}, \eqref{eq-plic-y4} and \eqref{eq-plic-cube-offset}, the argument of the square root may be negative before the case condition is tested. 
In this case -- since in the actual code, floating-point exception handling is turned off for performance reasons -- the resulting \texttt{NaN} of a square root of a negative number would not be captured in the case condition, leading to a wrong result.
An additional \texttt{fdim} function call in the square root solves this issue. 
In the implementation, we artificially exclude the edge case $x_4=0$ in order to instead of $\text{atan2}(y,x)$ use the faster $\text{atan}(y/x)$, giving the algorithm a $15\%$ speedup. In case branching would be undesirable, bit masking is also an option, but bit masking turned out to be slower even on GPUs.

Two edge cases still need to be taken into careful consideration: $n_1=0$ (2D) and $n_3=1$ (1D). $n_1=0$ restricts $\vec{n}$ to be in a 2D plane of two coordinate system axes and $n_3=1$ restricts $\vec{n}$ to be parallel to one of the coordinate system axes.
For the (1D) case, $n_3=1$ and the solution is always (5), so $n_1=n_2=0$ and $\min(n_1+n_2, n_3)=0$, simplifying equation \eqref{eq-plic-cube-offset} without loss of generality to
\begin{equation} \label{eq-plic-cube-offset-1D}
d=
\begin{cases}
V &\text{ (5) if }0\leq d\leq1
\end{cases}
\end{equation}
A clean derivation of the 1D case yields
\begin{equation}
d=V
\end{equation}
without any conditions, so it is a necessary requirement that both additional conditions in eq. \eqref{eq-plic-cube-offset-1D} must be fulfilled automatically. $d$ is in the range $0\leq d\leq\frac{n_1+n_2+n_3}{2}$, so here in the special case we have $0\leq d\leq\frac{0+0+1}{2}=\frac{1}{2}\leq 1$, which means $0\leq d\leq 1$ is indeed fulfilled automatically.

In the (2D) case, $n_1=0$ ($n_2=0$ is excluded here since it is already covered in the (1D) case, so here $n_2>0$). Here only intersection cases (2) or (5) are possible, $0<n_2 \leq n_3\leq1$ and $\min(n_1+n_2, n_3)=n_2$, simplifying eq. \eqref{eq-plic-cube-offset} without loss of generality to
\begin{equation} \label{eq-plic-cube-offset-2D}
d=
\begin{cases}
\sqrt{2\,V\,n_2\,n_3} &\text{ (2) if }\ \ 0\leq d\leq n_2\\
V\,n_3+\frac{n_2}{2} &\text{ (5) if }n_2\leq d\leq n_3
\end{cases}
\end{equation}
A clean derivation of the 2D case yields
\begin{equation}
d=
\begin{cases}
\sqrt{2\,V\,n_2\,n_3} &\text{ (2) if }\ \ \ \ \ \ \ d\leq n_2\\
V\,n_3+\frac{n_2}{2} &\text{ (5) if }n_2\leq d
\end{cases}
\end{equation}
which has simpler conditions, so again it is a necessary requirement that both additional conditions in eq. \eqref{eq-plic-cube-offset-2D} must be fulfilled automatically. $d$ is in the range $0\leq d\leq\frac{n_1+n_2+n_3}{2}$, so here with the special conditions we have $0\leq d\leq\frac{0+n_2+n_3}{2}\leq\frac{n_3+n_3}{2}=n_3$, meaning that both $0\leq d$ and $d\leq n_3$ are indeed fulfilled automatically.\\\\
In the above chosen $(5)\to(2)\to(1)\to(3)\to(4)$ implementation order, the 1D and 2D special cases are already covered in (5) and (2) at the beginning, so they both are excluded in the remaining intersection cases (1), (3) and (4), meaning that there $n_1,n_2,n_3>0$ are always given, resulting in $x_3>0$ in equation \eqref{eq-plic-x3}.

Listing \ref{list-plic-implementation} shows the fully optimized OpenCL C implementation of the analytic PLIC solution with equation \eqref{eq-plic-cube-offset-a-priori}.
\begin{lstlisting}[caption={Fully optimized OpenCL C implementation of our analytic PLIC solution.},captionpos=b,label={list-plic-implementation}]
float plic_cube_reduced(const float V, const float n1, const float n2, const float n3) {
	const float n1pn2=n1+n2, n3xV=n3*V;
	if(n1pn2<=2.0f*n3xV) return n3xV+0.5f*n1pn2; // case (5)
	const float V6n2n3=6.0f*n2*n3xV, sqn1=sq(n1);
	if(V6n2n3>=sq(n1) && 3.0f*n2*(2.0f*n3xV+n1-n2)<=sqn1) return 0.5f*n1+0.28867513f*sqrt(24.0f*n2*n3xV-sqn1); // case (2)
	if(V6n2n3<sqn1) return cbrt(V6n2n3*n1); // case (1)
	const float n1xn2=n1*n2;
	const float x3 = 81.0f*n1xn2*(n1pn2-2.0f*n3xV); // x3>0
	const float y32 = fdim(23328.0f*cb(n1xn2), sq(x3)); // y3>=0
	const float u3 = cbrt(sq(x3)+y32);
	const float d3 = n1pn2-(7.5595264f*n1xn2+0.26456684f*u3)*rsqrt(u3)*sin(0.5235988f-0.33333334f*atan(sqrt(y32)/x3)); // x3>0
	if(d3<=n3) return d3; // case (3)
	const float t4 = 9.0f*sq(n1pn2+n3)-18.0f;
	const float x4 = fmax(n1xn2*n3*(324.0f-648.0f*V), 1.1754944E-38f); // avoid edge case V==0.5 to make x4>0
	const float y42 = 4.0f*cb(t4)-sq(x4); // y4>=0
	const float u4 = cbrt(sq(x4)+y42);
	const float d4 = 0.5f*(n1pn2+n3)-(0.20998684f*t4+0.13228342f*u4)*rsqrt(u4)*sin(0.5235988f-0.33333334f*atan(sqrt(y42)/x4)); // x4>0
	return d4; // case (4)
}
float plic_cube(const float V0, const float3 n) { // unit cube - plane intersection: volume V0 in [0,1], normal vector n -> plane offset d0
	const float ax=fabs(n.x), ay=fabs(n.y), az=fabs(n.z), V=0.5f-fabs(V0-0.5f); // eliminate symmetry cases
	const float n1 = fmin(fmin(ax, ay), az);
	const float n3 = fmax(fmax(ax, ay), az);
	const float n2 = ax-n1+ay+az-n3;
	const float d = plic_cube_reduced(V, n1, n2, n3); // calculate PLIC with reduced symmetry
	return copysign(0.5f*(n1+n2+n3)-d, V0-0.5f); // apply symmetry for V0>0.5
}
\end{lstlisting}

\subsection{The analytic SZ solution}
The analytic PLIC solution by Scardovelli and Zaleski from 2000 \cite{scardovelli2000analytical} has been implemented in Fortran in 2016 by Kawano \cite{kawano2016simple} where it is used as comparison to the approximative  APPLIC method. 
Here we focus on the exact SZ solution.
The SZ solution is particularly interesting in that it builds upon the $L_1$-normalized plane normal vector instead of the more common $L_2$ normalization as used in our own solution in section \ref{sec-plic}. 
We first translate the Fortran implementation to OpenCL C, make it compatible with an $L_2$-normalized plane normal vector as input and rescale the result from the original $[0,1]$ to $[-\frac{|n_x|+|n_y|+|n_z|}{2},\frac{|n_x|+|n_y|+|n_z|}{2}]$. 
The implementation is provided in listing \ref{list-plic-implementation-sz}.

\begin{lstlisting}[caption={The Fortran implementation \cite{kawano2016simple} of the analytic SZ PLIC solution \cite{scardovelli2000analytical} translated to OpenCL C without further optimization.},captionpos=b,label={list-plic-implementation-sz}]
float plic_cube(const float V0, const float3 n) { // unit cube - plane intersection: volume V0 in [0,1], normal vector n -> plane offset d0
	const float l = fabs(n.x)+fabs(n.y)+fabs(n.z); // length in L1 norm
	const float ax=fabs(n.x)/l, ay=fabs(n.y)/l, az=fabs(n.z)/l, w=0.5f-fabs(V0-0.5f); // eliminate symmetry cases
	const float vm1 = fmin(fmin(ax, ay), az);
	const float vm3 = fmax(fmax(ax, ay), az);
	const float vm2 = fdim(1.0f, vm1+vm3); // ensure vm2>=0
	const float vm12 = vm1+vm2;
	float alpha = 0.0f;
	const float v1 = sq(vm1)/(6.0f*vm2*vm3+1E-25f);
	const float w6 = 6.0f*vm1*vm2*vm3*w;
	if(w<v1) {
		alpha = cbrt(w6); // case (1)
	} else if(w<v1+0.5f*(vm2-vm1)/vm3) {
		alpha = 0.5f*(vm1+sqrt(sq(vm1)+8.0f*vm2*vm3*(w-v1))); // case (2)
	} else {
		float v3;
		if(vm3<vm12) {
			v3 = (sq(vm3)*(3.0f*vm12-vm3)+sq(vm1)*(vm1-3.0f*vm3)+sq(vm2)*(vm2-3.0f*vm3))/(6.0f*vm1*vm2*vm3);
		} else {
			v3 = 0.5f*vm12/vm3;
			if(v3<=w) alpha = vm3*w+0.5f*vm12; // case (5)
		}
		if(alpha==0.0f) {
			float a0, a1, a2;
			if(w<v3) { // case (3)
				a2 = -3.0f*vm12;
				a1 = 3.0f*(sq(vm1)+sq(vm2));
				a0 = w6-cb(vm1)-cb(vm2);
			} else { // case (4)
				a2 = -1.5f;
				a1 = 1.5f*(sq(vm1)+sq(vm2)+sq(vm3));
				a0 = 0.5f*(w6-cb(vm1)-cb(vm2)-cb(vm3));
			}
			const float q0 = 0.16666667f*(a1*a2-3.0f*a0)-3.7037037E-2f*cb(a2); // 3.7037037E-2f = 1/27
			const float sp = sqrt(0.11111111f*sq(a2)-0.33333334f*a1);
			alpha = 2.0f*sp*cos(4.1887902f+0.33333334f*acos(q0/cb(sp)))-0.33333334f*a2; // 4.1887902f = 4/3*pi
		}
	}
	return l*copysign(0.5f-alpha, V0-0.5f); // rescale result and apply symmetry for V0>0.5
}
\end{lstlisting}
When closely studying the implementation in listing \ref{list-plic-implementation-sz}, we notice that the 1D edge case is poorly handled despite the addition of a tiny constant to the denominator for the calculation of \texttt{v1}. 
In the edge cases of the normal vector $\vec{n}=(1,0,0)^\mathrm{T}$ and the volume $V_0\in\{0,1\}$, the algorithm returns \texttt{-NaN} both in our OpenCL and in the original Fortran implementation.
By checking cases (5) and (2) first and by avoiding divisions by \texttt{vm1} and \texttt{vm2} before the checks for cases (5) and (2) respectively, we improve handling of the 1D edge case. We further apply some micro-optimization to significantly reduce the number of arithmetic operations, resulting in an optimized implementation of the SZ solution shown in listing \ref{list-plic-implementation-sz-optimized} below.
\begin{lstlisting}[caption={Fully optimized OpenCL C implementation of the SZ PLIC solution.},captionpos=b,label={list-plic-implementation-sz-optimized}]
float plic_cube_reduced(const float V, const float n1, const float n2, const float n3) { // optimized solution from SZ and Kawano
	const float n12=n1+n2, n3V=n3*V;
	if(n12<=2.0f*n3V) return n3V+0.5f*n12; // case (5)
	const float sqn1=sq(n1), n26=6.0f*n2, v1=sqn1/n26; // after case (5) check n2>0 is true
	if(v1<=n3V && n3V<v1+0.5f*(n2-n1)) return 0.5f*(n1+sqrt(sqn1+8.0f*n2*(n3V-v1))); // case (2)
	const float V6 = n1*n26*n3V;
	if(n3V<v1) return cbrt(V6); // case (1)
	const float v3 = n3<n12 ? (sq(n3)*(3.0f*n12-n3)+sqn1*(n1-3.0f*n3)+sq(n2)*(n2-3.0f*n3))/(n1*n26) : 0.5f*n12; // after case (2) check n1>0 is true
	const float sqn12=sqn1+sq(n2), V6cbn12=V6-cb(n1)-cb(n2);
	const bool case34 = n3V<v3; // true: case (3), false: case (4)
	const float a = case34 ? V6cbn12 : 0.5f*(V6cbn12-cb(n3));
	const float b = case34 ?   sqn12 : 0.5f*(sqn12+sq(n3));
	const float c = case34 ?     n12 : 0.5f;
	const float t = sqrt(sq(c)-b);
	return c-2.0f*t*sin(0.33333334f*asin((cb(c)-0.5f*a-1.5f*b*c)/cb(t)));
}
float plic_cube(const float V0, const float3 n) { // unit cube - plane intersection: volume V0 in [0,1], normal vector n -> plane offset d0
	const float ax=fabs(n.x), ay=fabs(n.y), az=fabs(n.z), V=0.5f-fabs(V0-0.5f), l=ax+ay+az; // eliminate symmetry cases, normalize n using L1 norm
	const float n1 = fmin(fmin(ax, ay), az)/l;
	const float n3 = fmax(fmax(ax, ay), az)/l;
	const float n2 = fdim(1.0f, n1+n3); // ensure n2>=0
	const float d = plic_cube_reduced(V, n1, n2, n3); // calculate PLIC with reduced symmetry
	return l*copysign(0.5f-d, V0-0.5f); // rescale result and apply symmetry for V0>0.5
}
\end{lstlisting}

\subsection{Iterative Solutions}
For comparison, we also provide iterative solutions using nested intervals in listing \ref{list-plic-implementation-ni} and Newton-Raphson in listing \ref{list-plic-implementation-nr}.
\begin{lstlisting}[caption={An OpenCL C implementation of the iterative nested intervals solution for cases (3) and (4).},captionpos=b,label={list-plic-implementation-ni}]
int log2_fast(const float x) { // evil log2 hack: log2(x)=(as_uint(x)>>23)-127
	return (as_uint(x)>>23)-127;
}
float plic_cube_reduced(const float V, const float n1, const float n2, const float n3) {
	const float n1pn2=n1+n2, n3xV=n3*V;
	if(n1pn2<=2.0f*n3xV) return n3xV+0.5f*n1pn2; // case (5)
	const float V6n2n3=6.0f*n2*n3xV, sqn1=sq(n1);
	if(V6n2n3>=sq(n1) && 3.0f*n2*(2.0f*n3xV+n1-n2)<=sqn1) return 0.5f*n1+0.28867513f*sqrt(24.0f*n2*n3xV-sqn1); // case (2)
	if(V6n2n3<sqn1) return cbrt(V6n2n3*n1); // case (1)
	const float V6n1n2n3 = V6n2n3*n1;
	float dmin, dmax, d;
	uint k;
	dmin=n2; dmax=n1+n2; d=0.5f*(dmin+dmax);
	k = (uint)log2_fast((dmax-dmin)*1.67772162E7f); // deterdmine number of interval halvings to reach machine precision
	for(uint i=0; i<=k; i++) {
		if(cb(d)-cb(d-n1)-cb(d-n2)<V6n1n2n3) dmin = d;
		else dmax = d;
		d = 0.5f*(dmin+dmax);
	}
	if(d<=n3) return d; // case (3)
	dmin=n3; dmax=0.5f*(n1+n2+n3); d=0.5f*(dmin+dmax);
	k = (uint)log2_fast((dmax-dmin)*1.67772162E7f); // deterdmine number of interval halvings to reach machine precision
	for(uint i=0; i<=k; i++) {
		if(cb(d)-cb(d-n1)-cb(d-n2)-cb(d-n3)<V6n1n2n3) dmin = d;
		else dmax = d;
		d = 0.5f*(dmin+dmax);
	}
	return d;
}
float plic_cube(const float V0, const float3 n) {
	const float n1 = fmin(fmin(fabs(n.x), fabs(n.y)), fabs(n.z)); // eliminate most cases due to symmetry
	const float n3 = fmax(fmax(fabs(n.x), fabs(n.y)), fabs(n.z));
	const float n2 = fabs(n.x)-n1+fabs(n.y)+fabs(n.z)-n3;
	const float V = 0.5f-fabs(V0-0.5f);
	const float d = plic_cube_reduced(V, n1, n2, n3);
	return copysign(0.5f*(n1+n2+n3)-d, V0-0.5f); // apply symmetry for V0>0.5
}
\end{lstlisting}
\begin{lstlisting}[caption={OpenCL C implementation of the iterative Newton-Raphson solution for cases (1), (3) and (4). Calculating case (1) with Newton-Raphson as well instead of the \texttt{cbrt()} function results in a very small, but noticeable improvement in performance when executed on the CPU with AVX2 vectorization.},captionpos=b,label={list-plic-implementation-nr}]
float plic_cube_reduced(const float V, const float n1, const float n2, const float n3) {
	const float n1pn2=n1+n2, n3xV=n3*V;
	if(n1pn2<=2.0f*n3xV) return n3xV+0.5f*n1pn2; // case (5)
	const float V6n2n3=6.0f*n2*n3xV, sqn1=sq(n1);
	if(V6n2n3>=sq(n1) && 3.0f*n2*(2.0f*n3xV+n1-n2)<=sqn1) return 0.5f*n1+0.28867513f*sqrt(24.0f*n2*n3xV-sqn1); // case (2)
	const float V6n1n2n3 = V6n2n3*n1;
	float dmin, dmax, d;
	if(V6n2n3<sqn1) {
		dmin=0.0f; dmax=n1; d=0.5f*(dmin+dmax);
		for(uint i=0; i<7; i++) {
			const float f = cb(d)-V6n1n2n3;
			const float fs = 3.0f*sq(d);
			d -= f/fs;
		}
		return d; // case (1)
	}
	dmin=n2; dmax=n1+n2; d=0.5f*(dmin+dmax);
	for(uint i=0; i<4; i++) {
		const float f = cb(d)-cb(d-n1)-cb(d-n2)-V6n1n2n3;
		const float fs = 3.0f*(sq(d)-sq(d-n1)-sq(d-n2));
		d -= f/fs;
	}
	if(d<=n3) return d; // case (3)
	dmin=n3; dmax=0.5f*(n1+n2+n3); d=0.5f*(dmin+dmax);
	for(uint i=0; i<4; i++) {
		const float f = cb(d)-cb(d-n1)-cb(d-n2)-cb(d-n3)-V6n1n2n3;
		const float fs = 3.0f*(sq(d)-sq(d-n1)-sq(d-n2)-sq(d-n3));
		d -= f/fs;
	}
	return d; // case (4)
}
float plic_cube(const float V0, const float3 n) {
	const float n1 = fmin(fmin(fabs(n.x), fabs(n.y)), fabs(n.z)); // eliminate most cases due to symmetry
	const float n3 = fmax(fmax(fabs(n.x), fabs(n.y)), fabs(n.z));
	const float n2 = fabs(n.x)-n1+fabs(n.y)+fabs(n.z)-n3;
	const float V = 0.5f-fabs(V0-0.5f);
	const float d = plic_cube_reduced(V, n1, n2, n3);
	return copysign(0.5f*(n1+n2+n3)-d, V0-0.5f); // apply symmetry for V0>0.5
}
\end{lstlisting}

\subsection{Performance and Accuracy Comparison}
Apart from floating-point errors, the SZ solution in listing \ref{list-plic-implementation-sz-optimized} and our own solution in listing \ref{list-plic-implementation} produce identical results.
For accuracy comparison, we define the error as
\begin{equation}
E_i:=|\text{plic\_cube\_inverse}(\text{plic\_cube}(V_0,\,\vec{n}_i),\,\vec{n}_i)-V_0|
\end{equation}
with $\text{plic\_cube\_inverse}(d_0,\,\vec{n}_i)$ referring to the implementation in listing \ref{list-inverse-plic-implementation} and $\text{plic\_cube}(V_0,\,\vec{n}_i)$ referring to the various PLIC implementations. 
We define the average error as follows:
\begin{align}
E_\text{avg}&:=\frac{1}{N\,L}\sum_{i=0}^{N\,L-1}(E_i)
\end{align}
The execution time for a 'blank run' 
\begin{equation}
E_{i,\text{blank}}:=|\text{plic\_cube\_inverse}(V_0-\frac{1}{2},\,\vec{n}_i)-V_0|
\end{equation}
containing the time for memory loads and stores as well as compute time for the inverse PLIC algorithm is measured and later subtracted from the execution times of the different PLIC variants in order to isolate their execution time. The time is also divided by the number of PLIC function evaluations $(N\,L)$ in order to obtain the time for a single execution.

\subsubsection{CPU Testing}
In this test, $\vec{n}_i$ is set to $N=4096$ different normal vectors, of which one is $(1,0,0)^\mathrm{T}$, one is $(\frac{1}{\sqrt{2}},\frac{1}{\sqrt{2}},0)^\mathrm{T}$, $510$ are random 2D directions in the $x$-$y$-plane and the remaining $3584$ are random 3D directions. For each of these, the volume $V_0$ is varied in the interval $[0,1]$ -- edge cases included -- in $L=4096$ equally spaced steps. The test is executed on a single core of a Coffee Lake Intel Core i7-8700K CPU at $4.0\,\text{GHz}$ AVX2 clock frequency with the MSVC C++ compiler and \texttt{/O2}, \texttt{/Oi}, \texttt{/Ot}, \texttt{/Qpar}, \texttt{/arch:AVX2}, \texttt{/fp:fast} and \texttt{/fp:except-} compiler flags set. 
The results are presented in table \ref{tab-perf-opt} below.\\
\begin{table*}[!htbp] \begin{center} \begin{tabular}{l|c c}
\hline
PLIC variant & execution time / $\text{ns}$ & $E_\text{avg}$ \\ 
\hline
\ref{list-plic-implementation-nr} Newton-Raphson        & $\,\ 16.2\pm1.9$ & $3.09\cdot10^{-8}$ \\
\ref{list-plic-implementation} our analytic solution    & $\,\ 42.9\pm2.5$ & $2.63\cdot10^{-8}$ \\
\ref{list-plic-implementation-sz-optimized} SZ solution optimized & $\,\ 43.7\pm2.4$ & $4.29\cdot10^{-8}$ \\
\ref{list-plic-implementation-sz} SZ solution by Kawano & $\,\ 46.0\pm2.0$ & $-\text{NaN}$ \\
\ref{list-plic-implementation-ni} nested intervals      & $270.1\pm3.2$ & $4.29\cdot10^{-8}$ \\
\hline
\end{tabular} \end{center}
\caption{Comparison of execution time and accuracy of the different PLIC variants with compiler optimizations enabled.} \label{tab-perf-opt}
\end{table*}\\
When AVX2 vectorization is available, Newton-Raphson is considerably faster than all the other solutions. However we note that our benchmark is a particularly synthetic scenario where vectorization is easily achievable for the compiler: The benchmark just calculates PLIC repeatedly in a loop. With FP32 and AVX2, every eight consecutive iterations are combined into one vectorized iteration, completing in far fewer clock cycles than the eight iterations separately would. In real-world applications, PLIC is not computed repeatedly in multiples of eight, greatly limiting vectorization potential such that the strict criteria for auto-vectorization \cite{microsoft2019vectorizer} might not be met. 
Surprisingly, our analytic solution and the optimized SZ solution are within margin of error regarding compute time. 

In the edge cases of the normal vector $\vec{n}=(1,0,0)^\mathrm{T}$ and the volume $V_0\in\{0,1\}$, the SZ solution by Kawano returns \texttt{-NaN}, which propagates through the entire error averaging procedure. For the IEEE-754 FP32 floating-point format, the machine epsilon is at $\epsilon=5.96\cdot10^{-8}$, meaning that for all other PLIC variants the average error $E_\text{avg}$ is within machine precision.\\
Without compiler optimization (\texttt{/Od}, \texttt{/Qpar-}), the execution time results are very different as shown in table \ref{tab-perf-dis}.
With AVX2 vectorization not available, Newton-Raphson considerably falls behind the analytic solutions.
\begin{table*}[!htbp] \begin{center} \begin{tabular}{l|c c}
\hline
PLIC variant & execution time / $\text{ns}$ & $E_\text{avg}$ \\ 
\hline
\ref{list-plic-implementation-sz-optimized} SZ solution optimized & $116.6\pm3.9$ & $4.70\cdot10^{-8}$ \\
\ref{list-plic-implementation} our analytic solution    & $124.1\pm2.3$ & $2.04\cdot10^{-8}$ \\
\ref{list-plic-implementation-sz} SZ solution by Kawano & $152.9\pm2.3$ & $-\text{NaN}$ \\
\ref{list-plic-implementation-nr} Newton-Raphson        & $180.7\pm3.4$ & $1.70\cdot10^{-8}$ \\
\ref{list-plic-implementation-ni} nested intervals      & $328.8\pm8.4$ & $2.10\cdot10^{-8}$ \\
\hline
\end{tabular} \end{center}
\caption{Comparison of execution time and accuracy of the different PLIC variants with compiler optimizations disabled.} \label{tab-perf-dis}
\end{table*}\\

\subsubsection{GPU Testing}
Since in many applications the target platform for the PLIC algorithm is the GPU, we also benchmark the different variants in OpenCL on an Nvidia Titan Xp GPU ($3840$ CUDA cores at $1582\,\text{MHz}$, driver version $442.59$, OpenCL $1.2$). 
The test here differs from the CPU C++ test in that for sufficient saturation of the parallel compute capabilities, the number of random normal vectors is increased to $67108864$, of which again one is $(1,0,0)^\mathrm{T}$, one is $(\frac{1}{\sqrt{2}},\frac{1}{\sqrt{2}},0)^\mathrm{T}$, $8388606$ are random 2D directions in the $x$-$y$-plane and the remaining $58720256$ are random 3D directions. 
For these normal vectors, PLIC is run in parallel and for each of them, the volume $V_0$ is varied in the interval $[0,1]$ -- edge cases included -- in $L=256$ equally spaced steps in series. 
For more accurate results through averaging, this test is run $64$ times and the mean execution time is averaged and divided by the number of parallel and serial PLIC computations, resulting in an average compute time of more than three magnitudes shorter than for single-core CPU execution as listed in table \ref{tab-perf-gpu} below. Even though the tests are designed differently for the CPU and GPU, this comparison is appropriate because in both cases the same algorithms are computed and the hardware is fully saturated. 
The table also includes the number of arithmetic operations (floating-point, integer and bit operations combined) $N_a$ and the number of branching operations $N_b$ in PTX assembly \cite{nvidia2019ptx}. 
GPUs are especially bad at branching, so a small $N_b$ is desired.
These operation counts -- in analogy to the compute time -- refer to the isolated PLIC variants with the background 'blank run' for memory loads and stores as well as the inverse PLIC algorithm subtracted.
\begin{table*}[!htbp] \begin{center} \begin{tabular}{l|c c c c}
\hline
PLIC variant & execution time / $\text{ps}$ & $N_a$ & $N_b$ & $E_\text{avg}$ \\ 
\hline
\ref{list-plic-implementation-sz-optimized} SZ solution optimized & $\,\ 12.8\pm1.0$ & $132$ & $12$ & $6.58\cdot10^{-8}$ \\
\ref{list-plic-implementation-sz} SZ solution by Kawano & $\,\ 16.0\pm1.5$ & $149$ & $14$ & $-\text{NaN}$ \\
\ref{list-plic-implementation} our analytic solution    & $\,\ 19.0\pm1.7$ & $189$ & $13$ & $5.73\cdot10^{-8}$ \\
\ref{list-plic-implementation-nr} Newton-Raphson        & $\,\ 19.2\pm1.8$ & $256$ & $11$ & $5.47\cdot10^{-8}$ \\
\ref{list-plic-implementation-ni} nested intervals      & $106.4\pm6.2$ & $106$ & $13$ & $2.86\cdot10^{-8}$ \\
\hline
\end{tabular} \end{center}
\caption{Comparison of execution time and accuracy of the different PLIC variants in OpenCL on a GPU.} \label{tab-perf-gpu}
\end{table*}
$N_a$ indicates that Newton-Raphson is unrolled by the compiler whereas nested intervals is not. 
The number of iterations to reach machine precision for nested intervals depends on the initial interval width, which is a function of the normal vector components. 
For Newton-Raphson, the number of branching operations $N_b$ is smallest, resulting in rather good performance. 
Nevertheless, our optimized implementation of the SZ solution pulls ahead rather significantly and thus is preferred by us.

\newpage
\section{Plane-Sphere Intersection} \label{sec-plic-sphere}
The PLIC problem can be extended to spherical cells with unit volume (see figure \ref{fig:plic-sphere}):
\begin{figure}[!htbp]
\centering \includegraphics[width=3.5cm]{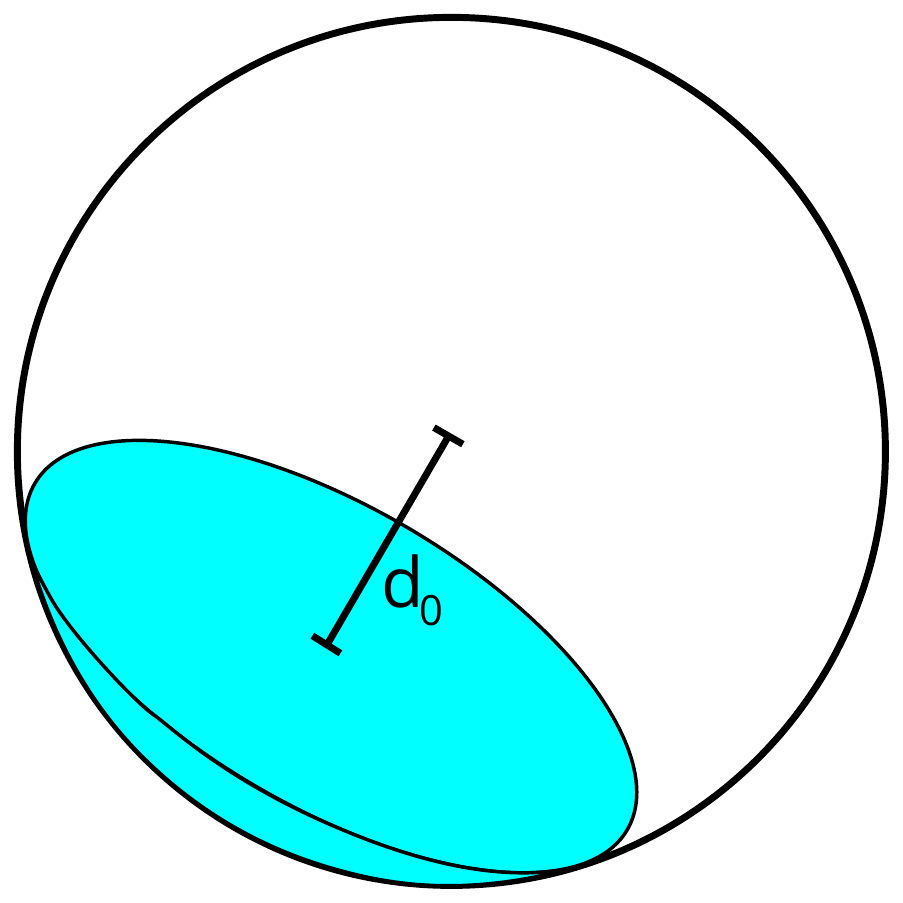}
\caption{Plane-sphere intersection. Here only one intersection case is possible.} 
\label{fig:plic-sphere}
\end{figure}
\begin{align}
1&=V=\frac{4}{3}\,\pi\,r^3\\
r&=\sqrt[3]{\frac{3}{4\,\pi}}
\end{align}
The offset along the plane normal vector is the desired result while the truncated volume is given. Since we have a sphere, the normal vector direction of the plane is irrelevant.
\begin{equation}
V_0\ \to\ d_0
\end{equation}
Calculating the inverse PLIC (getting the volume $V_0$ from a given offset $d_0\in[-r,r]$) is straight-forward and covered by the 'spherical cap' equation:
\begin{equation}
V_0=\frac{\pi}{3}\left(r+d_0\right)^2\left(2\,r-d_0\right)\in[0,1]
\end{equation}
Calculating the inverse function of the above equation again is quite difficult due to it being a third order polynomial with three complex solutions, but by using the same trick as in the plane-cube intersection case (equations \eqref{eq-plic-real} and \eqref{eq-plic-complex}), this real expression is obtained:
\begin{equation}
d_0=\sqrt[3]{\frac{6}{\pi}}\,\sin\left(\frac{\pi}{6}-\frac{1}{3}\,\atan2\left(2\sqrt{V_0-V_0^2},\ 2\,V_0-1\right)\right)
\end{equation}
Whilst the plane-sphere intersection solution is not particularly useful for VoF-LBM simulations, it might be of interest for some completely different applications.

\section{Application: Curvature Calculation for VoF-LBM on the GPU}

\subsection{Volume-of-Fluid Overview} \label{sec-vof-overview} \label{sec-curvature}
\begin{figure}[!htbp]
\begin{center}
\includegraphics[width=3.3cm]{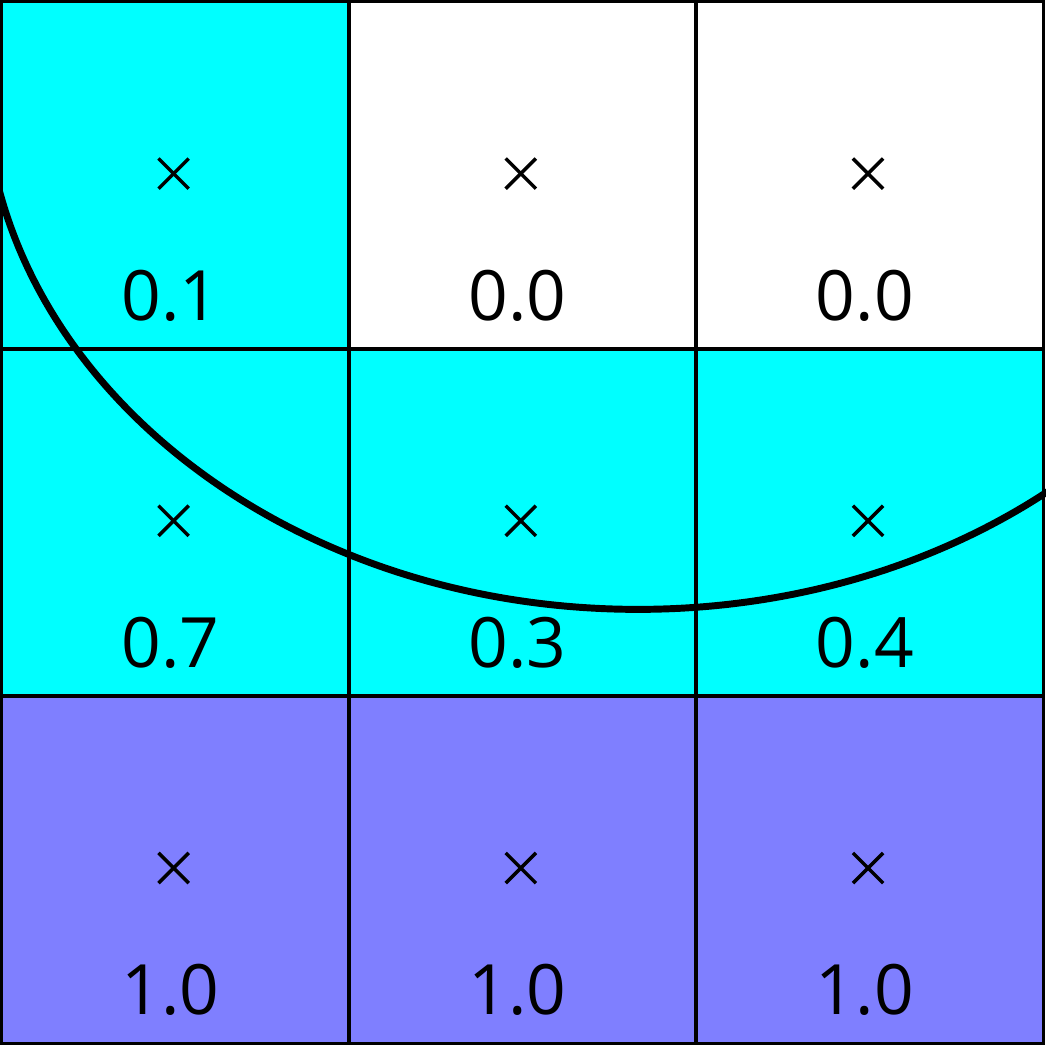}
\end{center}
\caption{The idea of the Volume-of-Fluid model illustrated in 2D: A sharp interface (black curved line) divides the \textit{gas} phase (white cells) from the \textit{fluid} phase (dark blue cells). 
Lattice points are located at the center of each cell. 
All cells through which the interface extends are called \textit{interface} cells (light blue). 
Every lattice cell has a fill level $\varphi\in[0,\,1]$ assigned to it, which is $\varphi=0$ for \textit{gas}, $\varphi=1$ for \textit{fluid} and $\varphi\in]0,\,1[$ for \textit{interface} -- based on where exactly the sharp interface cuts through.} \label{fig:vof-overview}
\end{figure}

Volume-of-Fluid (VoF) is a model to simulate a sharp, freely moving interface between a fluid and gas phase in a Cartesian lattice \cite{bogner2016curvature, korner2005lattice, thurey2005interactive, pohl2008high}. 
While it can be coupled to any flow solver, here we focus on its usage in conjunction with the Lattice-Boltzmann-method (LBM). 
The interface is ensured to be exactly one lattice cell thick at any time (illustrated in figure \ref{fig:vof-overview}). 
As an indicator for each lattice point type, the fill level $\varphi$ is introduced, whereby for \textit{fluid} lattice points $\varphi=1$, for \textit{interface} $1>\varphi>0$ and for \textit{gas} $\varphi=0$:
\begin{equation} \label{eq-vof-phi}
\varphi(\vec{x},t):=\frac{m(\vec{x},t)}{\rho(\vec{x},t)}
\begin{cases}
=1 &\text{if }\vec{x}\text{ is \textit{fluid}}\\
\in]0,\,1[ &\text{if }\vec{x}\text{ is \textit{interface}}\\
=0 &\text{if }\vec{x}\text{ is \textit{gas}}
\end{cases}
\end{equation}
Here $\rho$ is the density provided by the lattice Boltzmann method (LBM) and $m$ is the fluid mass. 
$m$ is a conserved quantity and cannot be gained or lost, only moved within the simulation box. 
\\
The key difficulty of modeling a free surface on a discretized lattice is to obtain the surface curvature, which is a necessary ingredient for calculating the surface tension via the Young-Laplace pressure
\begin{equation} \label{eq-young-laplace}
\Delta p=2\,\sigma\,\kappa
\end{equation}
with $\kappa=\frac{1}{R}$ denoting the local mean curvature and $\sigma$ denoting the surface tension parameter of the simulated fluid. 
The equation is easy in principle, but calculating $\kappa$ from the discretized interface geometry is not. 
Specifically, discretized interface here means that only a local $3^3$ neighborhood of fill levels $\varphi\in[0,1]$ is known in addition to the point in the center of this neighborhood being an \textit{interface} lattice point.
\begin{equation}
\varphi_0,\,...,\,\varphi_{26}\ \to\ \kappa
\end{equation}
The most common algorithm in literature \cite{bogner2016curvature, pohl2008high} is the curvature calculation via a least-squares paraboloid fit from a neighborhood of points located on the interface. 
It assumes the local interface to be a paraboloid, the specifics of which will be given in the following sections. 
Finding an appropriate set of neighboring points on the interface requires the PLIC solution.

\subsection{Obtaining neighboring interface points: PLIC point neighborhood} \label{sec-standard-plic}
\begin{figure*}[!htbp]
\begin{center}
  \begin{subfigure}{3.3cm}
    \includegraphics[width=\linewidth]{images/phi/phi_0.pdf}
    \caption{} \label{fig:a}
  \end{subfigure}%
  \hspace*{\fill} 
  \begin{subfigure}{3.3cm}
    \includegraphics[width=\linewidth]{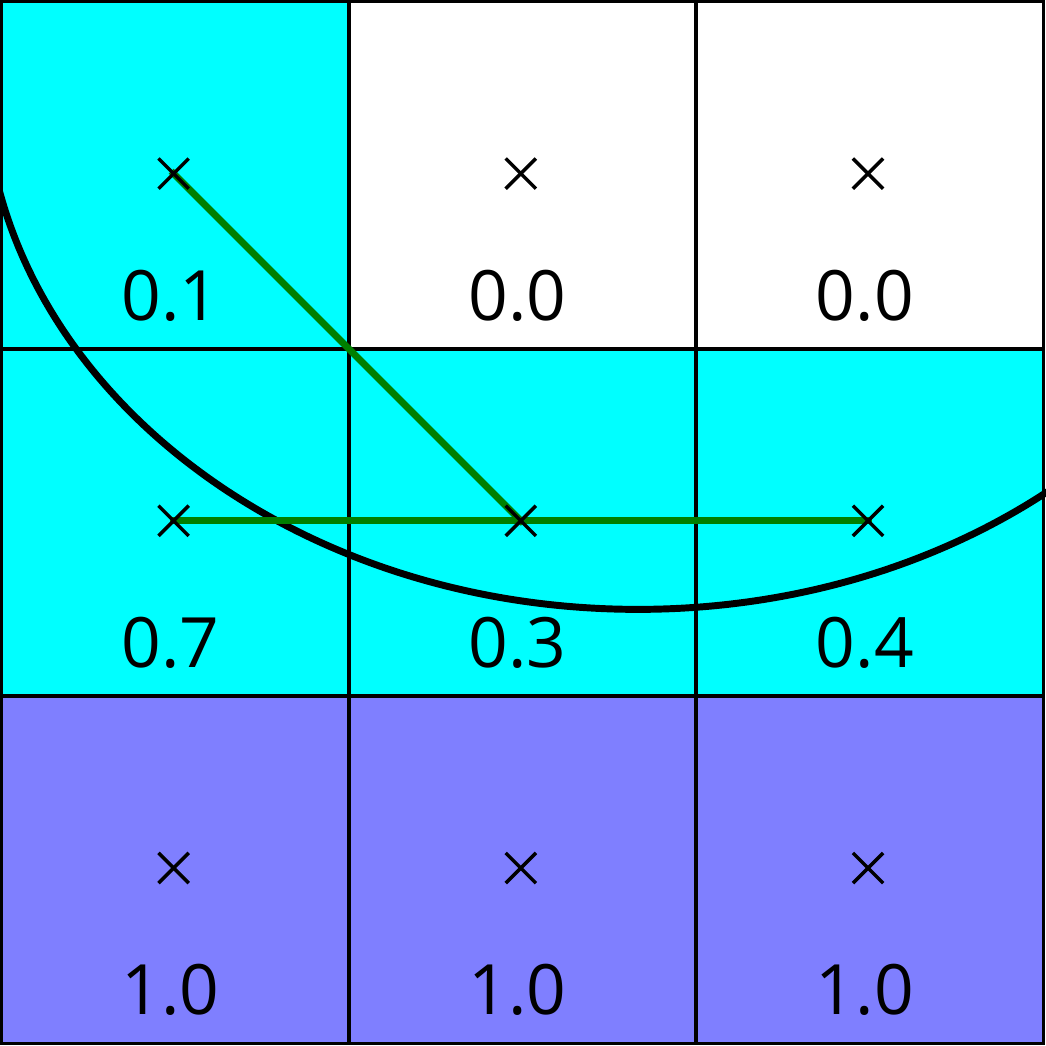}
    \caption{} \label{fig:b}
  \end{subfigure}%
  \hspace*{\fill} 
  \begin{subfigure}{3.3cm}
    \includegraphics[width=\linewidth]{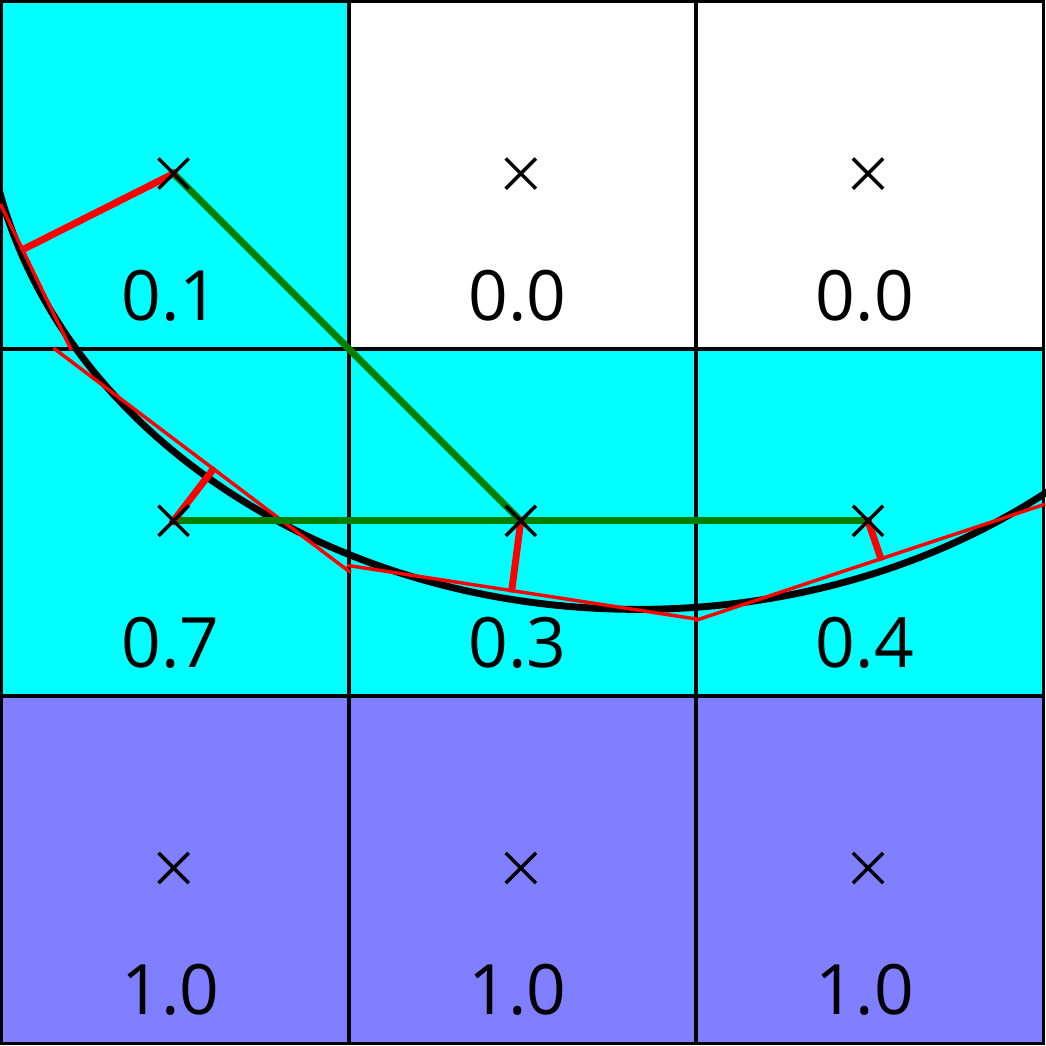}
    \caption{} \label{fig:c}
  \end{subfigure}%
  \hspace*{\fill} 
  \begin{subfigure}{3.3cm}
    \includegraphics[width=\linewidth]{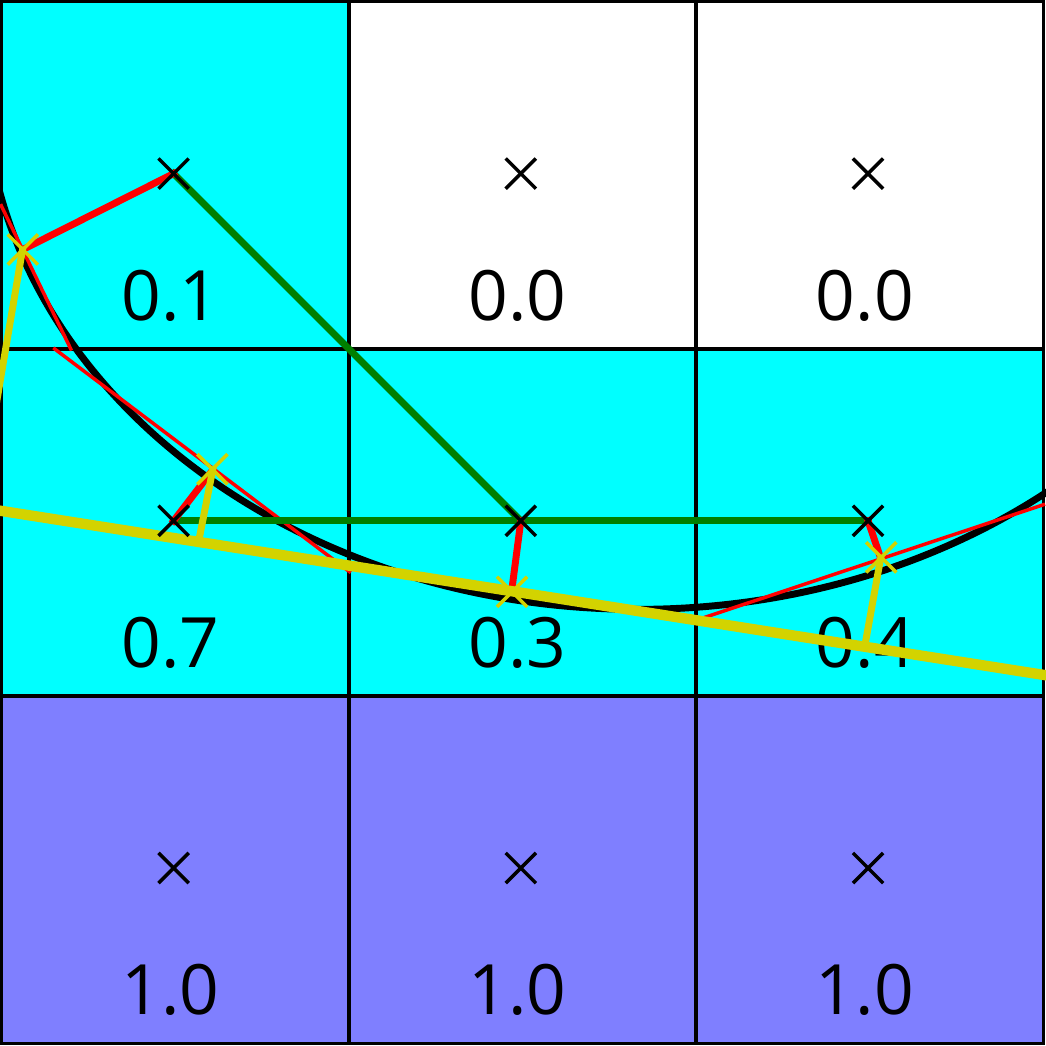}
    \caption{} \label{fig:d}
  \end{subfigure}%
  \hspace*{\fill} 
  \begin{subfigure}{3.3cm}
    \includegraphics[width=\linewidth]{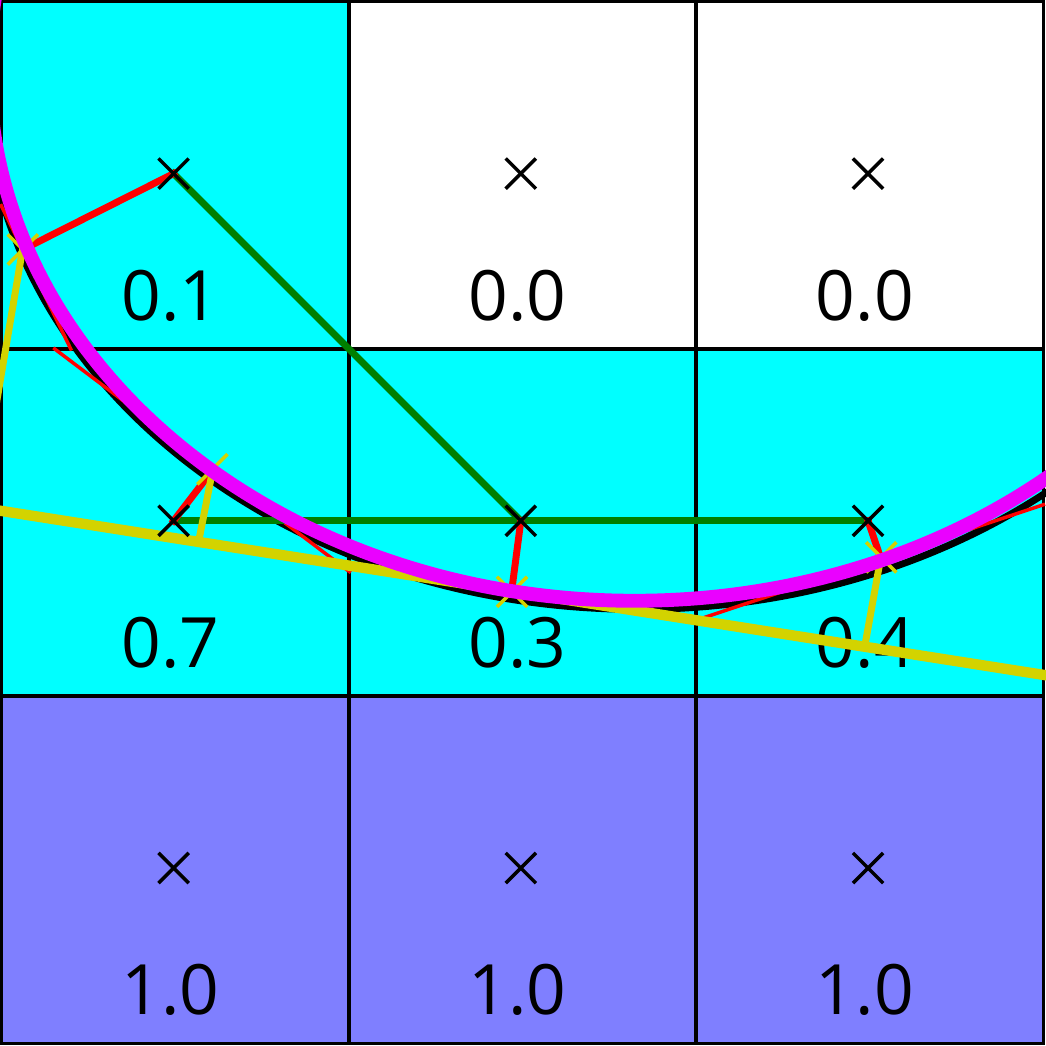}
    \caption{} \label{fig:e}
  \end{subfigure}%
\end{center}
\caption{The curvature calculation procedure with PLIC illustrated in 2D. 
(a) The fill levels of the interface lattice points indicate the position of the true interface (not known; here illustrated as a black curve), but only a $3^3$-neighborhood of these fill levels is available in memory. 
For obtaining the local curvature, the steps are to (b) identify all \textit{interface} neighbors (eq. \eqref{eq-vof-phi}), (c) correct the relative interface neighbor positions with the PLIC offset (section \ref{sec-plic-cube}), (d) rotate/translate these now PLIC-corrected points into a coordinate system (eq. \eqref{eq-curvature-coordinate-transformation}) with the PLIC-corrected center point being the origin and the z-axis being colinear with the local surface normal (appendix \ref{sec-curvature-normal}) and finally (e) perform a paraboloid fit with these points (appendix \ref{sec-least-squares-fit}).} \label{fig:curvature-fit}
\end{figure*}
Piecewise linear interface construction works on a $3^3$ neighborhood of an interface lattice point (illustrated in 2D in figure \ref{fig:a}). 
Within this neighborhood, only interface points other than the center interface point 
-- which is always interface -- are considered as candidates for the later fitting procedure (figure \ref{fig:b}). 
The difficult part is to accurately obtain the heights $z_i$ of at least five neighboring points located on the true interface (figure \ref{fig:c}).
For this, first the normal vector $\vec{n}$ of the center interface point is calculated via the Parker-Youngs approximation as described in the appendix \ref{sec-curvature-normal} in eq. \eqref{eq-curvature-normal-py}. 
A new coordinate system is introduced with its first base vector $\vec{b}_z$ defined as this normal vector. 
Then, the cross product with an arbitrary vector such as
\begin{equation}
\vec{r}:=(0.56270900,\,0.32704452,\,0.75921047)^ \mathrm{T}
\end{equation}
which is always non-colinear with $\vec{b}_z$ just by random chance is calculated to provide second and third orthonormal vectors
\begin{align}
\vec{b}_z&:=\vec{n}\\
\vec{b}_y&:=\frac{\vec{b}_z\cross\vec{r}}{|\vec{b}_z\cross\vec{r}|}\\
\vec{b}_x&:=\vec{b}_y\cross\vec{b}_z
\end{align}
forming the new coordinate system in which the z-axis is colinear with the surface normal and the center interface point is in the origin. Now the relative positions $\vec{e}_i$ (equal to the D3Q27 LBM streaming directions, eq. \eqref{eq-lbm-streaming-velocities}) of all neighboring interface lattice points are gathered and transformed into the rotated coordinate system. 
During this step, the approximate interface position of neighboring interface points (streaming directions, figure \ref{fig:b}) is corrected to the much more accurate interface position via the PLIC plane-cube intersection solution (section \ref{sec-plic-cube}, figure \ref{fig:c} and \ref{fig:d}):
\begin{equation} \label{eq-curvature-coordinate-transformation}
\vec{p}_i=\begin{pmatrix}
x_i\\
y_i\\
z_i
\end{pmatrix}:=\begin{pmatrix}
\vec{e}_i\scalar\vec{b}_x\\
\vec{e}_i\scalar\vec{b}_y\\
\vec{e}_i\scalar\vec{b}_z+d_0(\varphi_i,\,\vec{n})-d_0(\varphi_0,\,\vec{n})
\end{pmatrix}
\end{equation}
Here $i$ is only the subset of $\{0,..,26\}$ for which $0<\varphi_i<1$ is true (interface points). 
$d_0(V_0=\varphi_i,\,\vec{n})$ denotes the PLIC function (equation \eqref{eq-plic-d0}). Note that $d_0(V_0=\varphi_0,\,\vec{n})$ only needs to be calculated once while $d_0(V_0=\varphi_i,\,\vec{n})$ has to be calculated for each neighboring interface point and that the normal vectors of neighboring interface lattice points are approximated to be equal to the normal vector of the center lattice point. In theory, going with the separately calculated neighbor normal vectors -- which would require either an additional data buffer for normal vectors in memory or alternatively a $5^3$ neighborhood which would break the multi-GPU capabilities of the code -- should be more accurate, but our practical tests indicated no significant difference (see figure \ref{fig:curvature-error}).

The set of points $\vec{p}_i$ is then used to fit a local paraboloid.
This paraboloid (figure \ref{fig:e}) here lacks a vertical offset parameter as that is handled already by the center point being defined as the origin, reducing computational cost to a LU-decomposition of dimensionality $N=5$. The paraboloid has the form
\begin{equation}
z(x,y)=Ax^2+By^2+Cxy+Hx+Iy=:\vec{x}\scalar\vec{Q}
\end{equation}
with
\begin{align}
\vec{x}&:=(A,\,B,\,C,\,H,\,I)^ \mathrm{T}\\
\vec{Q}&:=(x^2,\,y^2,\,x\,y,\,x,\,y)^ \mathrm{T}
\end{align}
The solution vector $\vec{x}$ and thus the fitting parameters are calculated following the procedure in appendix \ref{sec-least-squares-fit}. 
Finally, the constants $A$, $B$, $C$, $H$ and $I$ are inserted into the analytic equation for the curvature \eqref{eq-paraboloid-curvature}, completing the algorithm.
\subsection{Validating Curvature Calculation}
\begin{figure}[!htbp]
\centering \includegraphics[width=8cm]{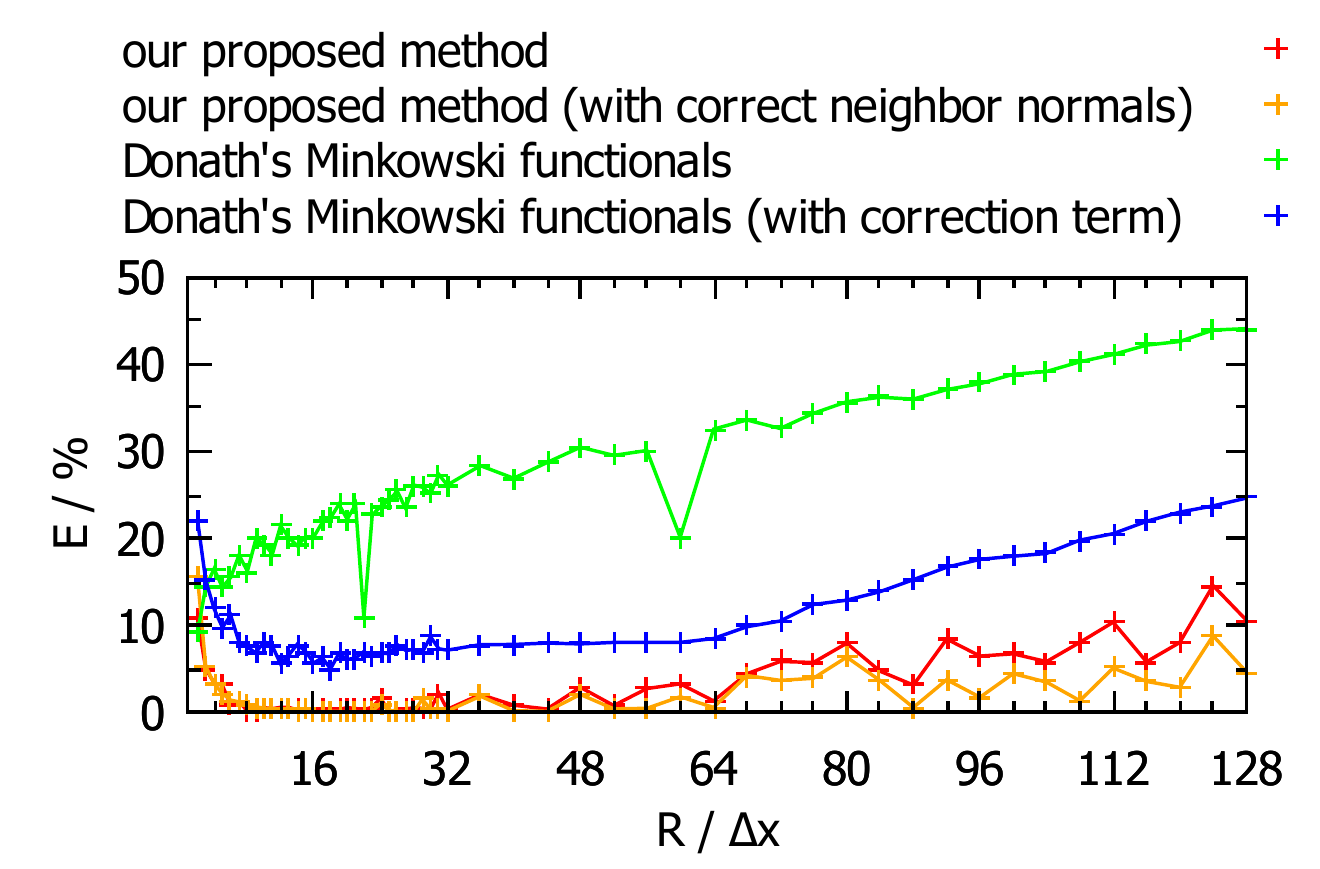}
\caption{$L_1$ error for our proposed curvature calculation method (red) depending on sphere radius. We also show what happens when we do the PLIC reconstruction for the \textit{interface} neighborhood not with the center normal vector, but with the normal vectors of the interface neighbors as discussed in section \ref{sec-standard-plic} (orange). For comparison, we provide the curvature approximation method proposed by Donath \cite{donath2011wetting} in both its variants (green and blue).} 
\label{fig:curvature-error}
\end{figure}

To test the accuracy of the presented curvature calculation method, we validate it on spherical drops of different radius $R=\frac{1}{\kappa_\text{theo}}$. The fill levels $\varphi$ are initialized with the inverse PLIC algorithm. To let the drop relax and the error converge, we simulate up to 50000 LBM time steps with D3Q19 SRT ($\tau=1$, $\sigma=0.001$). The curvature error is calculated using the $L_1$ error norm with summation over all \textit{interface} points:
\begin{equation}
E(\kappa):=\frac{\sum|\kappa_\text{sim}-\kappa_\text{theo}|}{\sum\kappa_\text{theo}}
\end{equation}
We plot the error in figure \ref{fig:curvature-error}. For low drop radius $R<32$, the error of our proposed method (red curve) is around $0.5\%$. However as $R$ is increased, the error also generally increases as a result of more \textit{interface} points in the summation and increased likelihood that the deviation in curvature for some points is much larger than average.\\
If we do not make the assumption that the PLIC normal direction of neighbor \textit{interface} points has to equal the center normal vector (orange curve), then accuracy is slightly improved. However the additional computational complexity may not justify this small improvement in accuracy.\\
For comparison, we also plot the curvature approximation method proposed by Donath \cite{donath2011wetting} without and with the proposed $\frac{\pi}{4}$ correction factor. As expected, the error is much larger, but this method is also computationally less expensive than ours and certainly has its use-cases as well.

\subsection{Application Example: Simulating a terminal Velocity Raindrop Impact}
\begin{figure*}[!htbp]
\begin{center}
  \begin{subfigure}{5.3cm}
    \includegraphics[width=\linewidth]{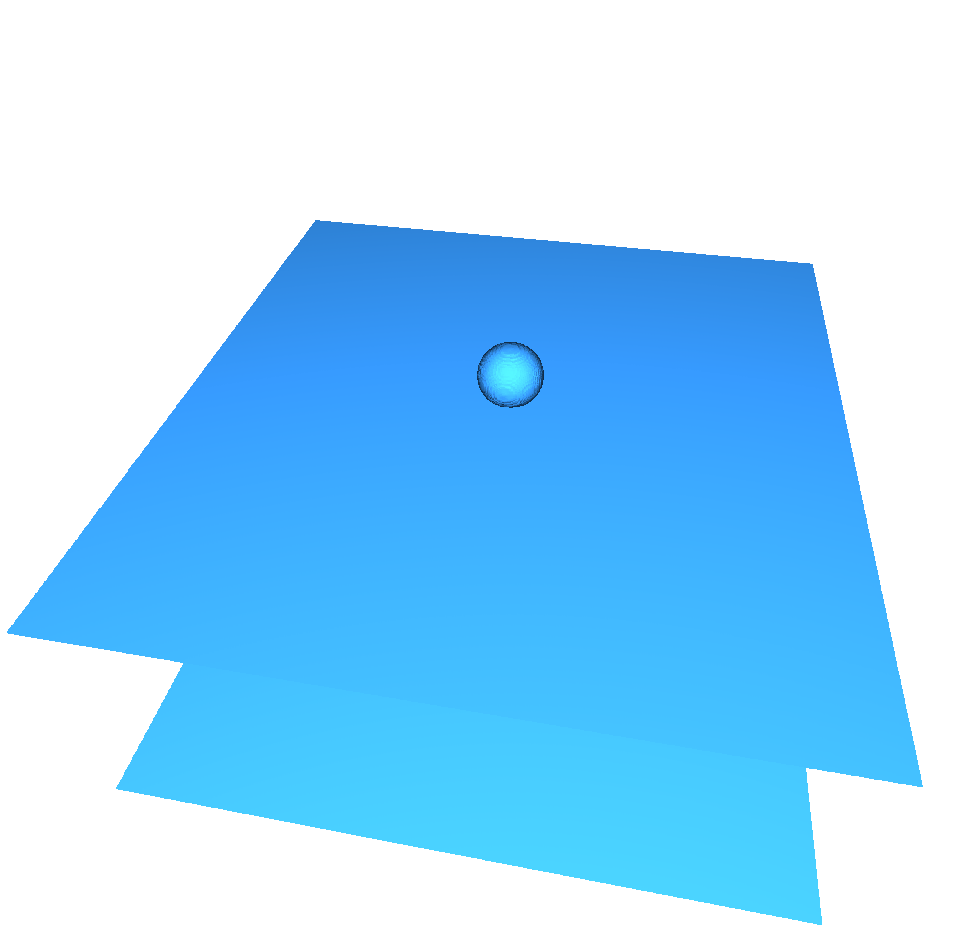}
    \caption{$t=0\,\text{ms}$} \label{fig:0ms}
  \end{subfigure}%
  \hspace*{\fill} 
  \begin{subfigure}{5.3cm}
    \includegraphics[width=\linewidth]{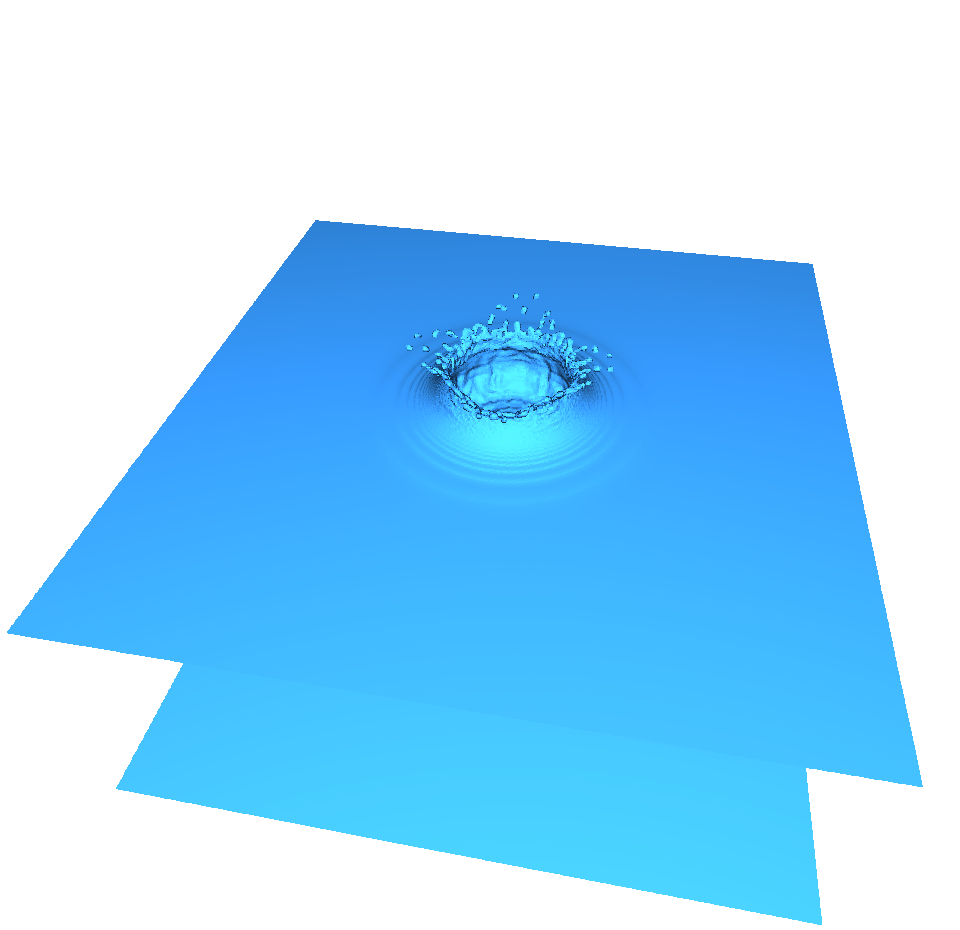}
    \caption{$t=1\,\text{ms}$} \label{fig:1ms}
  \end{subfigure}%
  \hspace*{\fill} 
  \begin{subfigure}{5.3cm}
    \includegraphics[width=\linewidth]{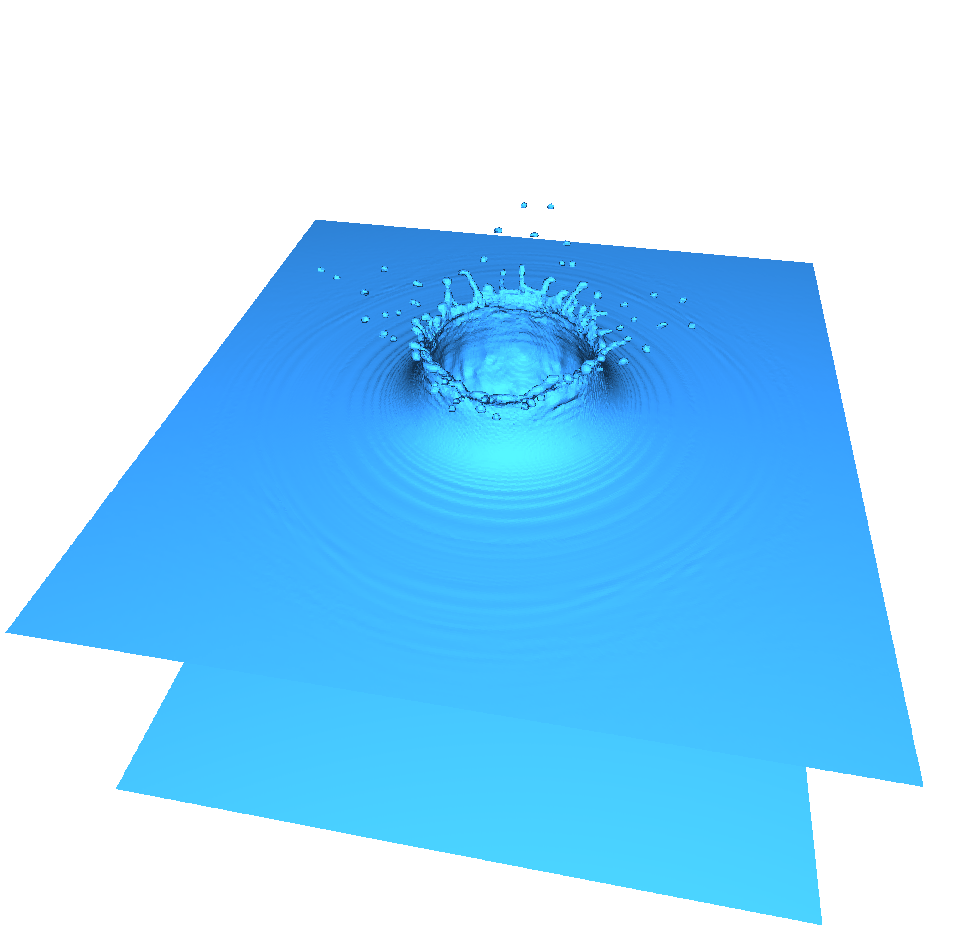}
    \caption{$t=2\,\text{ms}$} \label{fig:2ms}
  \end{subfigure}%
  \\ 
  \begin{subfigure}{5.3cm}
    \includegraphics[width=\linewidth]{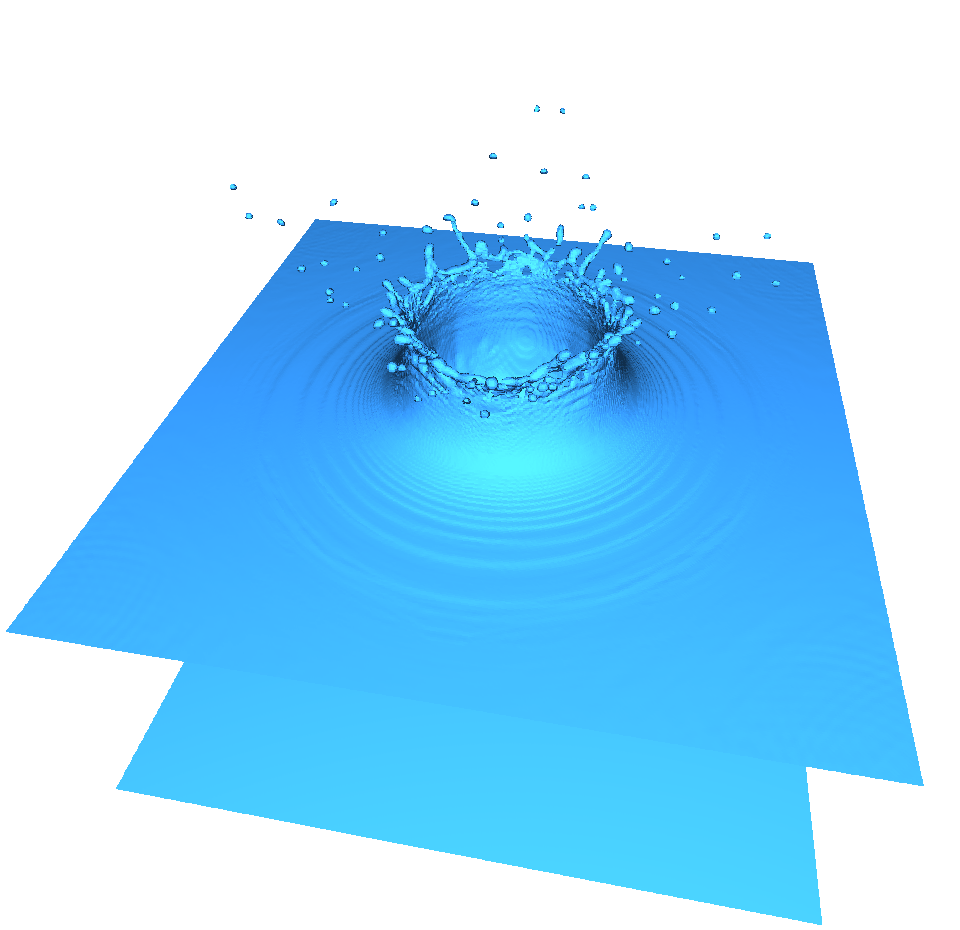}
    \caption{$t=3\,\text{ms}$} \label{fig:3ms}
  \end{subfigure}%
  \hspace*{\fill} 
  \begin{subfigure}{5.3cm}
    \includegraphics[width=\linewidth]{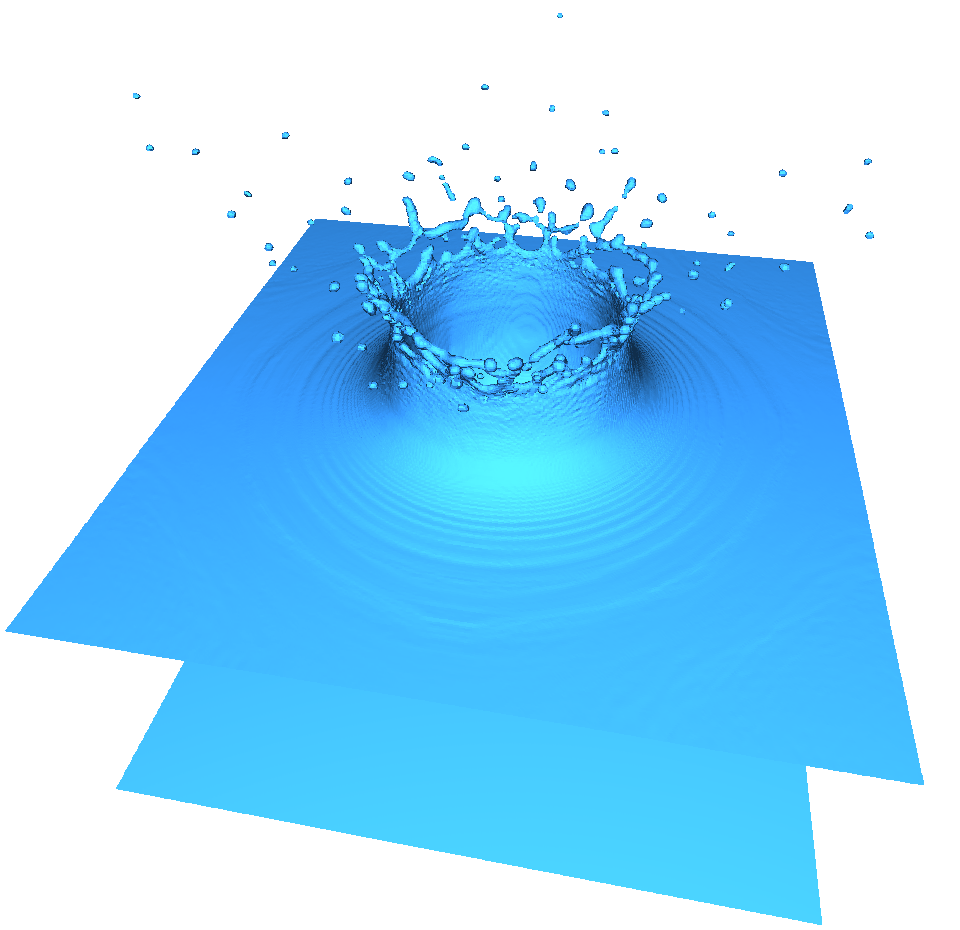}
    \caption{$t=4\,\text{ms}$} \label{fig:4ms}
  \end{subfigure}%
  \hspace*{\fill} 
  \begin{subfigure}{5.3cm}
    \includegraphics[width=\linewidth]{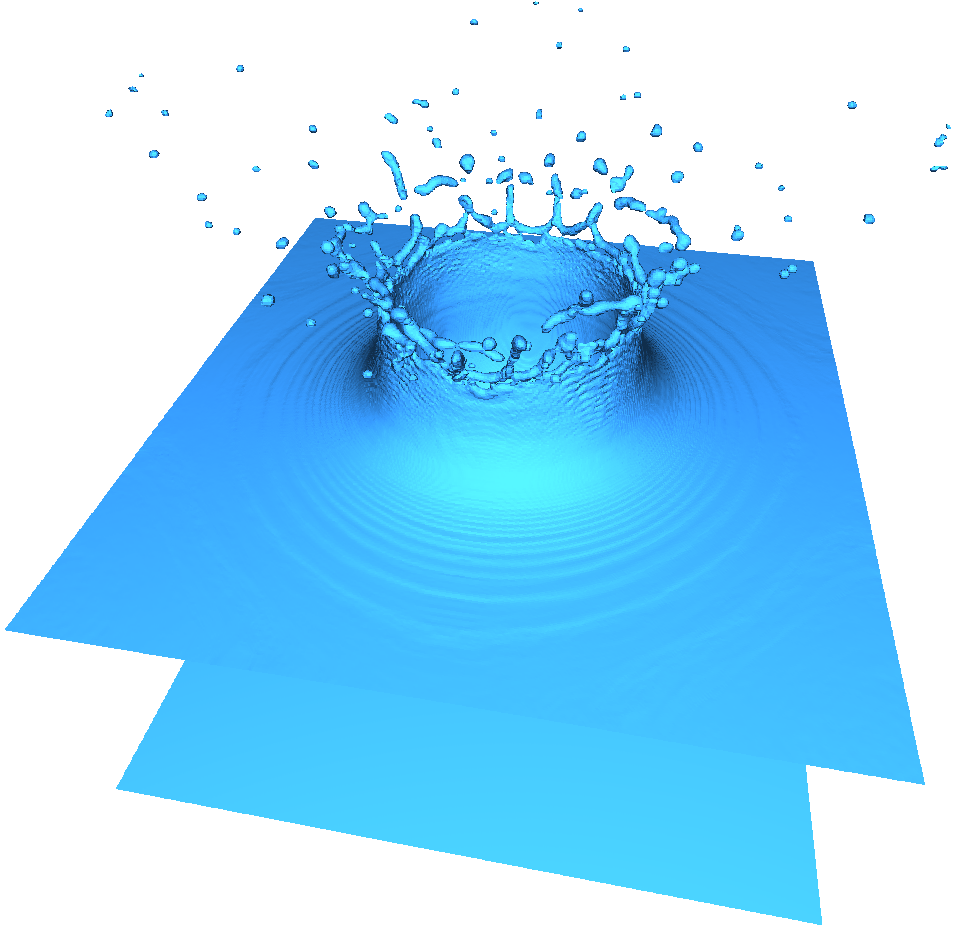}
    \caption{$t=5\,\text{ms}$} \label{fig:5ms}
  \end{subfigure}%
\end{center}
\caption{A $5\,\text{mm}$ diameter raindrop impacting a lake at $9.2\,\frac{\text{m}}{\text{s}}$ mean sea level pressure terminal velocity \cite{porcu2013effects} and $10\degree\text{C}$ water temperature simulated with the VoF-LBM GPU simulation code \textsl{FluidX3D} at a lattice resolution of $400\times400\times340$ with the D3Q19 discretization and the SRT collision operator. The simulation box in SI-units has the dimensions $5.00\,\text{cm}\times5.00\,\text{cm}\times4.25\,\text{cm}$ and the pool height is $2.00\,\text{cm}$. Compute time for this simulation is less than one minute on a single Nvidia Titan Xp GPU. Visualization is done with a custom GPU implementation of the marching cubes algorithm \cite{bourke1994polygonising, lorensen1987marching, vega2019fast}.} \label{fig:drop}
\end{figure*}

The analytic plane-cube intersection solution presented in this work has originally been developed for the VoF-LBM GPU simulation code \textsl{FluidX3D}, where we could observe a significant speedup compared to when an iterative nested-intervals solution is used. To illustrate this particular application of PLIC, we show a simulation of a $5\,\text{mm}$ diameter raindrop impact at terminal velocity in figure \ref{fig:drop}. The parameters for this simulation are $\textit{Re}=35195$, $\textit{We}=5702$, $\textit{Fr}=41.54$, $\textit{Ca}=0.1620$, $\textit{Bo}=3.3042$. The simulation code \textsl{FluidX3D} is documented in great detail in \cite{lehmann2019high}.

\section{Conclusions}
We derived an analytic solution to the PLIC problem and compared it to the existing solution by Scardovelli and Zaleski \cite{scardovelli2000analytical} in two variants: an implementation by Kawano \cite{kawano2016simple} and an improved and micro-optimized version thereof. 
We furthermore compared these three analytic solutions to two iterative solutions using Newton-Raphson and nested intervals. 
We provide OpenCL C implementations of all variants as well as the inverse PLIC formulation.
We observed that in a synthetic benchmark scenario where AVX2 vectorization is available on the CPU, the Newton-Raphson solution (listing \ref{list-plic-implementation-nr}) is considerably faster than all other solutions because the analytic solutions require trigonometric functions which cannot be vectorized. 
Our benchmark calculated PLIC repeatedly in a loop. With FP32 and AVX2, every eight consecutive iterations are combined into one vectorized iteration, completing in far fewer clock cycles than the eight iterations separately would. In real-world applications however, PLIC is not computed repeatedly in multiples of eight, greatly limiting vectorization potential.
Without vectorization, on both the CPU and GPU the analytic solutions are faster, with our micro-optimized version of the SZ solution as presented in listing \ref{list-plic-implementation-sz-optimized} being fastest. 
For a generic PLIC problem, this is the solution we recommend.
In the most common application of PLIC -- curvature calculation for Volume-of-Fluid LBM, which we presented and also validated on spherical drops -- profiling revealed PLIC to be the main bottleneck regarding compute time. 
Here, our proposed fast PLIC solution led to significant speed up of VoF calculations.
We hope that our findings will also make other simulation codes more computationally efficient.


\newpage
\section{Acknowledgements}
We gratefully acknowledge computing time provided by the SuperMUC system of the Leibniz Rechenzentrum, Garching.
We further acknowledge support through the computational resources provided by the Bavarian Polymer Institute.
We acknowledge the NVIDIA Corporation for donating a Titan Xp GPU for our research.\\

\section{Declarations}
\subsection{Funding}
ML acknowledges funding by the Deutsche Forschungsgemeinschaft (DFG, German Research Foundation) - Project number 391977956  - SFB 1357 ''Microplastics'' (subproject B04).
\subsection{Conflicts of interest}
The authors declare no potential conflicts of interests.
\subsection{Availability of data and material}
All data is archived and available upon request.
\subsection{Code availability}
All code beyond what is provided in the listings is archived and available upon request.
\subsection{Authors' contributions}
Analytic calculations were done by ML (section \ref{sec-plic-cube}) and SG (section \ref{sec-plic-sphere}). ML did programming and testing. ML wrote the initial draft and created the illustrations. ML and SG did review and editing on the draft.

\section{References}
\printbibliography[heading=none]

@article{kruger2017lattice,
  title={The lattice Boltzmann method},
  author={Kr{\"u}ger, Timm and Kusumaatmaja, Halim and Kuzmin, Alexandr and Shardt, Orest and Silva, Goncalo and Viggen, Erlend Magnus},
  journal={Springer International Publishing},
  volume={10},
  pages={978--3},
  year={2017},
  publisher={Springer}
}

@book{chapman1990mathematical,
  title={The mathematical theory of non-uniform gases: an account of the kinetic theory of viscosity, thermal conduction and diffusion in gases},
  author={Chapman, Sydney and Cowling, Thomas George and Burnett, David},
  year={1990},
  publisher={Cambridge university press}
}

@inproceedings{purqon2017accuracy,
  title={Accuracy and Numerical Stabilty Analysis of Lattice Boltzmann Method with Multiple Relaxation Time for Incompressible Flows},
  author={Purqon, Acep and others},
  booktitle={Journal of Physics: Conference Series},
  volume={877},
  number={1},
  pages={012035},
  year={2017},
  organization={IOP Publishing}
}

@article{obrecht2011new,
  title={A new approach to the lattice Boltzmann method for graphics processing units},
  author={Obrecht, Christian and Kuznik, Fr{\'e}d{\'e}ric and Tourancheau, Bernard and Roux, Jean-Jacques},
  journal={Computers \& Mathematics with Applications},
  volume={61},
  number={12},
  pages={3628--3638},
  year={2011},
  publisher={Elsevier}
}

@article{wittmann2016hardware,
  title={Hardware-effiziente, hochparallele Implementierungen von Lattice-Boltzmann-Verfahren f{\"u}r komplexe Geometrien},
  author={Wittmann, Markus},
  year={2016}
}

@article{delbosc2014optimized,
  title={Optimized implementation of the Lattice Boltzmann Method on a graphics processing unit towards real-time fluid simulation},
  author={Delbosc, Nicolas and Summers, Jonathan L and Khan, AI and Kapur, Nik and Noakes, Cath J},
  journal={Computers \& Mathematics with Applications},
  volume={67},
  number={2},
  pages={462--475},
  year={2014},
  publisher={Elsevier}
}

@inproceedings{herschlag2018gpu,
  title={Gpu data access on complex geometries for d3q19 lattice boltzmann method},
  author={Herschlag, Gregory and Lee, Seyong and Vetter, Jeffrey S and Randles, Amanda},
  booktitle={2018 IEEE International Parallel and Distributed Processing Symposium (IPDPS)},
  pages={825--834},
  year={2018},
  organization={IEEE}
}

@article{mawson2014memory,
  title={Memory transfer optimization for a lattice Boltzmann solver on Kepler architecture nVidia GPUs},
  author={Mawson, Mark J and Revell, Alistair J},
  journal={Computer Physics Communications},
  volume={185},
  number={10},
  pages={2566--2574},
  year={2014},
  publisher={Elsevier}
}

@article{wittmann2013comparison,
  title={Comparison of different propagation steps for lattice Boltzmann methods},
  author={Wittmann, Markus and Zeiser, Thomas and Hager, Georg and Wellein, Gerhard},
  journal={Computers \& Mathematics with Applications},
  volume={65},
  number={6},
  pages={924--935},
  year={2013},
  publisher={Elsevier}
}

@article{kuznik2010lbm,
  title={LBM based flow simulation using GPU computing processor},
  author={Kuznik, Fr{\'e}d{\'e}ric and Obrecht, Christian and Rusaouen, Gilles and Roux, Jean-Jacques},
  journal={Computers \& Mathematics with Applications},
  volume={59},
  number={7},
  pages={2380--2392},
  year={2010},
  publisher={Elsevier}
}

@online{nvidia2019ptx,
  title = {Parallel Thread Execution ISA Version 6.4},
  author = {NVIDIA Corporation},
  year = 2019,
  url = {https://docs.nvidia.com/cuda/parallel-thread-execution/index.html},
  urldate = {2019-10-25}
}

@misc{haeusl2019mpi,
  title={MPI-based multi-GPU extension of the Lattice Boltzmann Method},
  author={Fabian Häusl},
  year={2019}
}

@book{pohl2008high,
  title={High performance simulation of free surface flows using the lattice Boltzmann method},
  author={Pohl, Thomas},
  year={2008},
  publisher={Verlag Dr. Hut}
}

@article{eberly2000least,
  title={Least squares fitting of data},
  author={Eberly, David},
  journal={Chapel Hill, NC: Magic Software},
  year={2000}
}

@article{korner2005lattice,
  title={Lattice Boltzmann model for free surface flow for modeling foaming},
  author={K{\"o}rner, Carolin and Thies, Michael and Hofmann, Torsten and Th{\"u}rey, Nils and R{\"u}de, Ulrich},
  journal={Journal of Statistical Physics},
  volume={121},
  number={1-2},
  pages={179--196},
  year={2005},
  publisher={Springer}
}

@article{thurey2005interactive,
  title={Interactive free surface fluids with the lattice Boltzmann method},
  author={Th{\"u}rey, Nils and K{\"o}rner, C and R{\"u}de, U},
  journal={Technical Report05-4. University of Erlangen-Nuremberg, Germany},
  year={2005}
}

@phdthesis{schreiber2010gpu,
  title={GPU based simulation and visualization of fluids with free surfaces},
  author={Schreiber, Martin and Neumann, DTMP},
  year={2010},
  school={Diploma Thesis, Technische Universit{\"a}t M{\"u}nchen}
}

@book{parker1992two,
  title={Two and three dimensional Eulerian simulation of fluid flow with material interfaces},
  author={Parker, BJ and Youngs, DL},
  year={1992},
  publisher={Atomic Weapons Establishment}
}

@book{pressley2010elementary,
  title={Elementary differential geometry},
  author={Pressley, Andrew N},
  year={2010},
  publisher={Springer Science \& Business Media}
}

@book{abbena2017modern,
  title={Modern differential geometry of curves and surfaces with Mathematica},
  author={Abbena, Elsa and Salamon, Simon and Gray, Alfred},
  year={2017},
  publisher={Chapman and Hall/CRC}
}

@inproceedings{yu2007focal,
  title={Focal surfaces of discrete geometry},
  author={Yu, Jingyi and Yin, Xiaotian and Gu, Xianfeng and McMillan, Leonard and Gortler, Steven},
  booktitle={ACM International Conference Proceeding Series},
  volume={257},
  pages={23--32},
  year={2007}
}

@article{har1995curvature,
  title={Curvature of curves and surfaces--a parabolic approach},
  author={Har’el, Zvi},
  journal={Department of Mathematics, Technion--Israel Institute of Technology},
  year={1995},
  publisher={Citeseer}
}

@article{jia2018gaussian,
  title={Gaussian and Mean Curvatures},
  author={Jia, Yan-Bin},
  year={2018}
}

@article{popinet2009accurate,
  title={An accurate adaptive solver for surface-tension-driven interfacial flows},
  author={Popinet, St{\'e}phane},
  journal={Journal of Computational Physics},
  volume={228},
  number={16},
  pages={5838--5866},
  year={2009},
  publisher={Elsevier}
}

@article{bogner2016curvature,
  title={Curvature estimation from a volume-of-fluid indicator function for the simulation of surface tension and wetting with a free-surface lattice Boltzmann method},
  author={Bogner, Simon and R{\"u}de, Ulrich and Harting, Jens},
  journal={Physical Review E},
  volume={93},
  number={4},
  pages={043302},
  year={2016},
  publisher={APS}
}

@article{janssen2013enhanced,
  title={On enhanced non-linear free surface flow simulations with a hybrid LBM--VOF model},
  author={Jan{\ss}en, Christian F and Grilli, Stephan T and Krafczyk, Manfred},
  journal={Computers \& Mathematics with Applications},
  volume={65},
  number={2},
  pages={211--229},
  year={2013},
  publisher={Elsevier}
}

@article{youngs1982time,
  title={Time-dependent multi-material flow with large fluid distortion},
  author={Youngs, David L},
  journal={Numerical methods for fluid dynamics},
  year={1982},
  publisher={Academic Press}
}

@misc{bourke1994polygonising,
  title={Polygonising a scalar field},
  author={Bourke, Paul},
  year={1994}
}

@article{lorensen1987marching,
  title={Marching cubes: A high resolution 3D surface construction algorithm},
  author={Lorensen, William E and Cline, Harvey E},
  journal={ACM siggraph computer graphics},
  volume={21},
  number={4},
  pages={163--169},
  year={1987},
  publisher={ACM New York, NY, USA}
}

@article{vega2019fast,
  title={A Fast and Memory-Saving Marching Cubes 33 Implementation with the Correct Interior Test},
  author={Vega, David and Abache, Javier and Coll, David},
  journal={Journal of Computer Graphics Techniques Vol},
  volume={8},
  number={3},
  year={2019}
}

@article{youngs1984interface,
  title={An interface tracking method for a 3D Eulerian hydrodynamics code},
  author={Youngs, David L},
  journal={Atomic Weapons Research Establishment (AWRE) Technical Report},
  volume={44},
  number={92},
  pages={35},
  year={1984}
}

@article{scardovelli2000analytical,
  title={Analytical relations connecting linear interfaces and volume fractions in rectangular grids},
  author={Scardovelli, Ruben and Zaleski, Stephane},
  journal={Journal of Computational Physics},
  volume={164},
  number={1},
  pages={228--237},
  year={2000},
  publisher={Elsevier}
}

@misc{lehmann2019high,
  title={High Performance Free Surface LBM on GPUs},
  author={Moritz Lehmann},
  year={2019}
}

@article{jafari2007improved,
  title={An improved three-dimensional model for interface pressure calculations in free-surface flows},
  author={Jafari, Ali and Shirani, Ebrahim and Ashgriz, Nasser},
  journal={International Journal of Computational Fluid Dynamics},
  volume={21},
  number={2},
  pages={87--97},
  year={2007},
  publisher={Taylor \& Francis}
}

@article{kawano2016simple,
  title={A simple volume-of-fluid reconstruction method for three-dimensional two-phase flows},
  author={Kawano, Akio},
  journal={Computers \& Fluids},
  volume={134},
  pages={130--145},
  year={2016},
  publisher={Elsevier}
}

@online{microsoft2019vectorizer,
  title = {Vectorizer and parallelizer messages},
  author = {Microsoft Corporation},
  year = 2019,
  url = {https://docs.microsoft.com/en-us/cpp/error-messages/tool-errors/vectorizer-and-parallelizer-messages?view=vs-2019},
  urldate = {2020-05-11}
}

@article{porcu2013effects,
  title={Effects of altitude on maximum raindrop size and fall velocity as limited by collisional breakup},
  author={Porc{\`u}, Federico and D’adderio, Leo Pio and Prodi, Franco and Caracciolo, Clelia},
  journal={Journal of the atmospheric sciences},
  volume={70},
  number={4},
  pages={1129--1134},
  year={2013}
}

@article{xing2007lattice,
  title={Lattice Boltzmann-based single-phase method for free surface tracking of droplet motions},
  author={Xing, Xiu Qing and Butler, David Lee and Yang, Chun},
  journal={International journal for numerical methods in fluids},
  volume={53},
  number={2},
  pages={333--351},
  year={2007},
  publisher={Wiley Online Library}
}

@article{donath2011verification,
  title={Verification of surface tension in the parallel free surface lattice Boltzmann method in waLBerla},
  author={Donath, Stefan and Mecke, Klaus and Rabha, Swapna and Buwa, Vivek and R{\"u}de, Ulrich},
  journal={Computers \& fluids},
  volume={45},
  number={1},
  pages={177--186},
  year={2011},
  publisher={Elsevier}
}

@book{donath2011wetting,
  title={Wetting Models for a Parallel High-performance Free Surface Lattice Boltzmann Method: Benetzungsmodelle F{\"u}r Eine Parallele Lattice-Boltzmann-Methode Mit Freien Oberfl{\"a}chen},
  author={Donath, Stefan},
  year={2011},
  publisher={Verlag Dr. Hut}
}

@article{anderl2014free,
  title={Free surface lattice Boltzmann with enhanced bubble model},
  author={Anderl, Daniela and Bogner, Simon and Rauh, Cornelia and R{\"u}de, Ulrich and Delgado, Antonio},
  journal={Computers \& Mathematics with Applications},
  volume={67},
  number={2},
  pages={331--339},
  year={2014},
  publisher={Elsevier}
}

\newpage
\section{Appendix A: PLIC Inversion with Mathematica}\label{appendix-a}
\vspace*{1cm}
\includegraphics[page=1,trim={1.8cm 1cm 2cm 2.5cm},clip,width=0.45\textwidth]{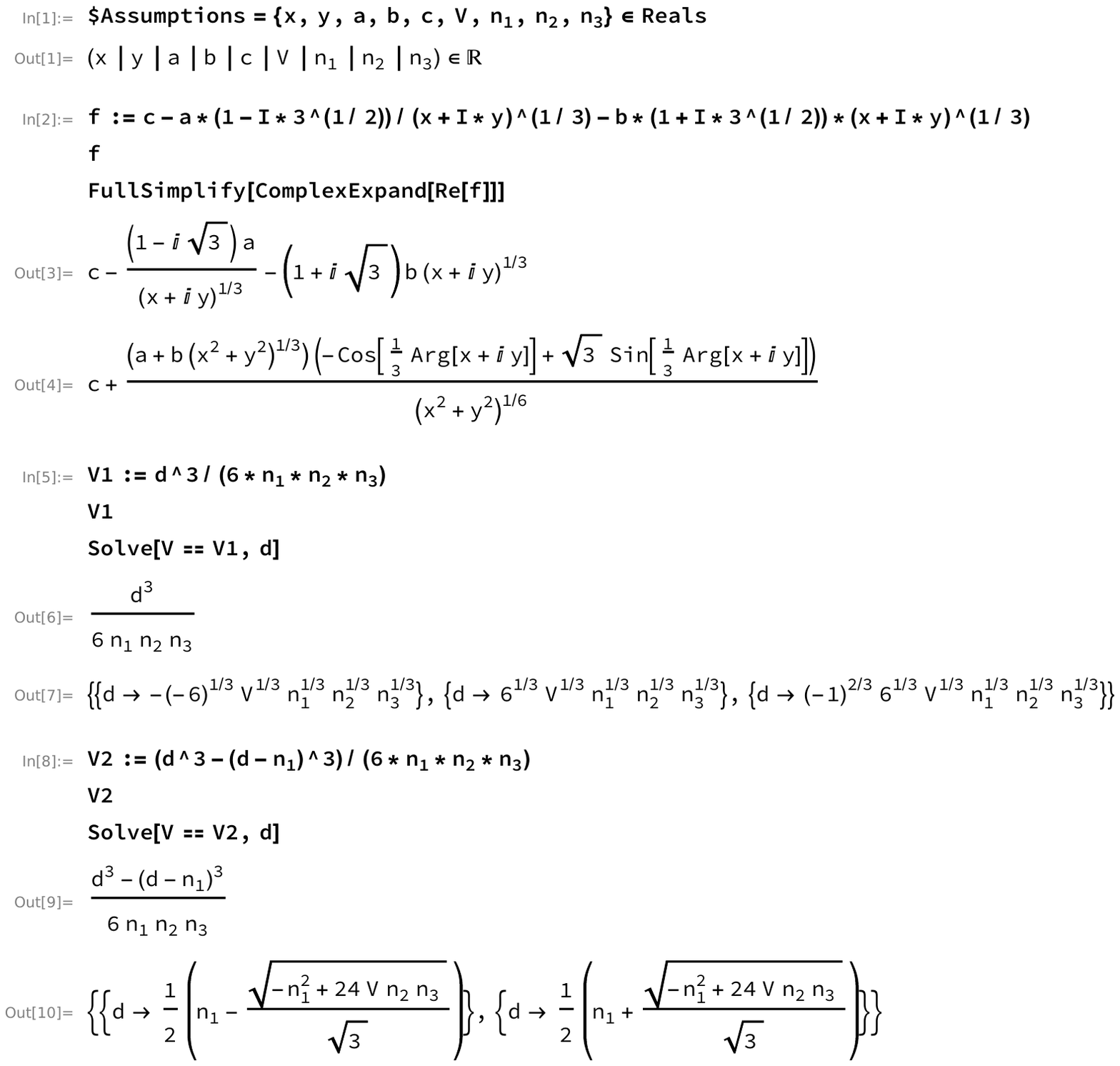}\\
\includegraphics[page=2,trim={1.8cm 6cm 2cm 2.5cm},clip,width=0.45\textwidth]{images/plic-solution.pdf}\\
\vspace*{2cm}
\includegraphics[page=3,trim={1.8cm 1cm 2cm 2.5cm},clip,width=0.45\textwidth]{images/plic-solution.pdf}\\
\includegraphics[page=4,trim={0.6cm 6cm 2cm 2.5cm},clip,width=0.45\textwidth]{images/plic-solution.pdf}\\


\newpage
\markedsection{Appendix B: Paraboloid Curvature, Interface Normal and Least-Squares Fit}{Appendix B}

\subsection{Calculating the Interface Normal Vector from a $3^3$ Neighborhood} \label{sec-curvature-normal}
\newpage\hfill\\ Calculating the normal vector on an interface lattice point in a $3^3$ neighborhood in which all fill levels $\varphi_i$ are known works by applying the gradient to the fill levels:
\begin{strip}
\begin{multline}
\nabla\varphi(x,y,z)=
\begin{pmatrix}
\frac{\partial}{\partial x}\varphi(x,y,z)\\
\frac{\partial}{\partial y}\varphi(x,y,z)\\
\frac{\partial}{\partial z}\varphi(x,y,z)
\end{pmatrix}\approx
\frac{1}{18}\begin{pmatrix}
\varphi_1+\varphi_7+\varphi_9+\varphi_{13}+\varphi_{15}+\varphi_{19}+\varphi_{21}+\varphi_{23}+\varphi_{26}\\
\varphi_3+\varphi_7+\varphi_{11}+\varphi_{14}+\varphi_{17}+\varphi_{19}+\varphi_{21}+\varphi_{24}+\varphi_{25}\\
\varphi_5+\varphi_9+\varphi_{11}+\varphi_{16}+\varphi_{18}+\varphi_{19}+\varphi_{22}+\varphi_{23}+\varphi_{25}
\end{pmatrix}-\\-\frac{1}{18}\begin{pmatrix}
\varphi_2+\varphi_8+\varphi_{10}+\varphi_{14}+\varphi_{16}+\varphi_{20}+\varphi_{22}+\varphi_{24}+\varphi_{25}\\
\varphi_4+\varphi_8+\varphi_{12}+\varphi_{13}+\varphi_{18}+\varphi_{20}+\varphi_{22}+\varphi_{23}+\varphi_{26}\\
\varphi_6+\varphi_{10}+\varphi_{12}+\varphi_{15}+\varphi_{17}+\varphi_{20}+\varphi_{21}+\varphi_{24}+\varphi_{26}
\end{pmatrix}=\frac{1}{18}\sum_{i=1}^{26}\vec{e}_i\,\varphi_i
\end{multline}
\end{strip}
This is called the center of mass (CM) method:
\begin{equation}
\vec{n}_\text{CM}:=-\frac{\sum_{i=1}^{26}\vec{e}_i\,\varphi_i}{|\sum_{i=1}^{26}\vec{e}_i\,\varphi_i|}
\end{equation}
$\vec{e}_i$ are the directions from the center point of the $3^3$-neighborhood to all of its 26 neighbors including itself:
\begin{strip}
\begin{equation} \label{eq-lbm-streaming-velocities}
\vec{e}_i=\begin{Bmatrix}
0&\pm1&   0&   0&\pm1&\pm1&   0&\pm1&\pm1&   0&\pm1&\pm1&\pm1&\mp1\\
0&   0&\pm1&   0&\pm1&   0&\pm1&\mp1&   0&\pm1&\pm1&\pm1&\mp1&\pm1\\
0&   0&   0&\pm1&   0&\pm1&\pm1&   0&\mp1&\mp1&\pm1&\mp1&\pm1&\pm1
\end{Bmatrix},\ \ \ \ i\in[0,26]
\end{equation}
\end{strip}
Another more accurate approach is the Parker-Youngs (PY) approximation \cite{parker1992two, donath2011wetting} which assigns different weights\footnote{We kindly note that \cite{pohl2008high} and \cite{lehmann2019high} provide the weights in the wrong order.} to the gradient components, similar to a Sobel filter:
\begin{equation} \label{eq-curvature-normal-py}
\vec{n}_\text{PY}:=-\frac{\sum_{i=1}^{26}w_i\,\vec{e}_i\,\varphi_i}{|\sum_{i=1}^{26}w_i\,\vec{e}_i\,\varphi_i|}
\end{equation}
with
\begin{equation}
w_i:=
\begin{cases}
\ 4&\text{ for }|\vec{c}_i|=\ \ 1\\
\ 2&\text{ for }|\vec{c}_i|=\sqrt{2}\\
\ 1&\text{ for }|\vec{c}_i|=\sqrt{3}
\end{cases}
\end{equation}
According to \cite{pohl2008high}, the average error for CM is approximately $4\degree$ while for PY it is approximately $1\degree$. For the surface curvature algorithms below, the more accurate and equally fast PY method is used.\\\\

\subsection{Analytic Curvature of a Paraboloid}
A paraboloid curve is described by
\begin{equation} \label{eq-paraboloid-equation}
z=f(x,y)=A\,x^2+B\,y^2+C\,x\,y+H\,x+I\,y+J
\end{equation}
where $A$, $B$, $C$, $H$, $I$ and $J$ are fitting parameters.
For such a 2D surface in 3D space in the Monge patch $x,\,y,\,z=f(x,y)$, the mean curvature \cite[p.185]{pressley2010elementary}\cite{abbena2017modern, yu2007focal, har1995curvature, jia2018gaussian} is 
\begin{equation} \label{eq-mean-curvature}
\kappa:=\frac{f_{xx}\left(f_y^2+1\right)+f_{yy}\left(f_x^2+1\right)-2\,f_{xy}\,f_x\,f_y}{2\left(\sqrt{f_x^2+f_y^2+1}\right)^3}
\end{equation}
The partial derivatives of eq. \eqref{eq-paraboloid-equation} evaluated at the point $(x=0,\,y=0)$ are
\begin{align} \label{eq-curvature-derivatives}
\at{f_{xx}}{x=y=0}&=2\,A\\
\at{f_{yy}}{x=y=0}&=2\,B\\
\at{f_{xy}}{x=y=0}&=C\\
\at{f_x}{x=y=0}&=\at{2\,A\,x+C\,y+H}{x=y=0}=H\\
\at{f_y}{x=y=0}&=\at{2\,B\,y+C\,x+\,I\,}{x=y=0}=\,I
\end{align}
so that the mean curvature for the paraboloid at the origin is given by
\begin{equation} \label{eq-paraboloid-curvature}
\kappa:=\frac{A\,(I^2+1)+B\,(H^2+1)-C\,H\,I}{\left(\sqrt{H^2+I^2+1}\right)^3}.
\end{equation}
We note here that \cite{bogner2016curvature} in equation (13) have an erroneous factor $2$ and that \cite{popinet2009accurate} use a different definition of the mean curvature. 
The strategy for finding the required fitting parameters is to apply a least-squares fit on a neighborhood of points on the interface.

\subsection{Curvature from Least-Squares Paraboloid Fit} \label{sec-least-squares-fit}
The least-squares method \cite{eberly2000least} is a procedure for fitting an analytic curve -- here a Monge patch -- to a set of discretized points located nearby the analytic curve. The general idea is to define the total error as a general expression of all fitting parameters and the entire set of discretized points and then find its global minimum by zeroing its gradient.\\
The analytic curve first needs to be written in a dot product form
\begin{equation}
z(x,y)=\vec{x}\scalar\vec{Q}
\end{equation}
with $\vec{x}$ being defined as the vector of parameters that define the curve and $\vec{Q}=\vec{Q}(x,y)$ being an expression of the continuous coordinates $x$ and $y$. This equation is then discretized to a set of individual data points $(x_i,\,y_i,\,z_i)$
\begin{equation}
z_i(x_i,y_i)\approx\vec{x}\scalar\vec{Q}_i
\end{equation}
with $\vec{Q}_i=\vec{Q}_i(x_i,y_i)$ being a vector containing expressions only dependent on a discretized set of points $(x_i,\,y_i)$ whose corresponding $z$-component $z_i$ is located close to the curve. In this notation, the error $E$ between the z-positions of the analytic curve $\vec{x}\scalar\vec{Q}$ and a set of z-positions of at least $N$ neighboring interface points $z_i$ is defined by summing up the squared differences
\begin{equation}
E(\vec{x})=\sum_{i=0}^{N}(\vec{x}\scalar\vec{Q}_i-z_i)^2
\end{equation}
whereby $N$ denotes the dimensionality which is equal to the number of desired fitting parameters. The gradient of the error $E$ is calculated and set to zero, where the error must have a global minimum:
\begin{equation}
\nabla E(\vec{x})=2\sum_{i=0}^{N}(\vec{x}\scalar\vec{Q}_i-z_i)\,\vec{Q}_i=0
\end{equation}
With some algebra, this equation is then transformed into a linear equation
\begin{align}
\left(\sum_{i=0}^{N}\vec{Q}_i\vec{Q_i}^ \mathrm{T}\right)\vec{x}&=\sum_{i=0}^{N}z_i\,\vec{Q}_i\\
\textbf{M}:=\sum_{i=0}^{N}\vec{Q}_i\vec{Q_i}^ \mathrm{T}\ &\ \ \ \ \ \ \ \ \vec{b}:=\sum_{i=0}^{N}z_i\,\vec{Q}_i\\
\textbf{M}\,\vec{x}&=\vec{b}
\end{align}
which is solved by LU-decomposition and provides the desired solution $\vec{x}$ that uniquely defines the curve.

Note that the matrix $\textbf{M}$ is always symmetrical, meaning that only the upper half and diagonal have to be calculated explicitly and the lower half is copied over. This reduces computational cost significantly due to every matrix element being a sum over an expression depending on all fitted points. In case there are less than $N$ data points available (lattice points next to solid boundaries may have less \textit{interface} neighbors), the regular fitting will not work. Instead, then the amount of fitting parameters is decreased to match the number of available data points by reducing dimensionality in the LU-decomposition. The ignored fitting parameters will remain zero. 

Finally, from the solution vector $\vec{x}$ the constants defining the fitted curve are extracted and the curvature is calculated from them using equation \eqref{eq-paraboloid-curvature}.

\end{document}